\documentclass[%
unsortedaddress,
 preprint,
 amsmath,amssymb,
 aip,
]{revtex4-2}
\usepackage{color}
\usepackage{graphicx}
\usepackage{dcolumn}
\usepackage{bm}
\usepackage{hyperref}

\usepackage[utf8]{inputenc}
\usepackage[T1]{fontenc}
\usepackage{mathptmx}
\usepackage{etoolbox}
\makeatletter
\def\@email#1#2{%
 \endgroup
 \patchcmd{\titleblock@produce}
 {\frontmatter@RRAPformat}
 {\frontmatter@RRAPformat{\produce@RRAP{*#1\href{mailto:#2}{#2}}}\frontmatter@RRAPformat}
 {}{}
}%
\makeatother

\begin{document}

\preprint{APS/123-QED}

\title{First-principle study of spin transport property in $L1_0$-FePd(001)/graphene heterojunction}

\author{Hayato Adachi}
\affiliation{Department of Electrical and Electronic Engineering, Graduate School of Engineering, Kobe University, 1-1 Rokkodai-cho, Nada-ku, Kobe 651-8501, Japan}

\author{Ryuusuke Endo}
\affiliation{Department of Electrical and Electronic Engineering, Graduate School of Engineering, Kobe University, 1-1 Rokkodai-cho, Nada-ku, Kobe 651-8501, Japan}

\author{Hikari Shinya}
\affiliation{
 Center for Spintronics Research Network (CSRN), University of Tokyo,
 7-3-1, Hongo, Bunkyo-ku, Tokyo 113-8656, Japan
}
\affiliation{
 Institute for Chemical Research, Kyoto University,
 Gokasho, Uji, Kyoto 611-0011, Japan
}
\affiliation{
 Center for Science and Innovation in Spintronics (CSIS), Tohoku University,
 2-1-1, Katahira, Aoba-ku, Miyagi 980-8577, Japan
}
\affiliation{
 Center for Spintronics Research Network (CSRN), Osaka University,
 1-3, Machikaneyama, Toyonaka, Osaka 560-8531, Japan
}

\author{Hiroshi Naganuma}
\affiliation{Center for Innovative Integrated Electronics Systems (CIES), Tohoku University, 468-1 Aramaki Aza Aoba, Aoba, Sendai, Miyagi, 980-8572, Japan}
\affiliation{Center for Spintronics Integrated Systems (CSIS), Tohoku University, 2-2-1 Katahira Aoba, Sendai, Miyagi 980-8577 Japan}
\affiliation{Center for Spintronics Research Network (CSRN), Tohoku University, 2-1-1 Katahira, Aoba, Sendai, Miyagi 980-8577 Japan}
\affiliation{Graduate School of Engineering, Tohoku University, 6-6-05, Aoba, Aoba-ku, Sendai, Miyagi, 980-8579, Japan}

\author{Tomoya Ono}
\affiliation{Department of Electrical and Electronic Engineering, Graduate School of Engineering, Kobe University, 1-1 Rokkodai-cho, Nada-ku, Kobe 651-8501, Japan}

\author{Mitsuharu Uemoto}
\email{uemoto@eedept.kobe-u.ac.jp}
\affiliation{Department of Electrical and Electronic Engineering, Graduate School of Engineering, Kobe University, 1-1 Rokkodai-cho, Nada-ku, Kobe 651-8501, Japan}

\date{\today}

\begin{abstract}
In our previous work, we synthesized a metal/2D material heterointerface consisting
 of $L1_0$-ordered iron--palladium (FePd) and graphene (Gr) called FePd(001)/Gr.
This system has been explored by both experimental measurements and theoretical calculations.
In this study, we focus on a heterojunction composed of FePd and multilayer graphene referred to as FePd(001)/$m$-Gr/FePd(001), where $m$ represents the number of graphene layers.
We perform first-principles calculations to predict their spin-dependent transport properties.
The quantitative calculations of spin-resolved conductance and magnetoresistance (MR) ratio ($150$ -- $200~\%$) suggest that the proposed structure can function as a magnetic tunnel junction in spintronics applications.
We also find that an increase in $m$ not only reduces conductance but also changes transport properties from the tunneling behavior to the graphite $\pi$-band-like behavior.
Additionally, we investigate the spin-transfer torque-induced magnetization switching behavior of our junction structures using micromagnetic simulations.
Furthermore, we examine the impact of lateral displacements (``sliding'') at the interface and find that the spin transport properties remain robust despite these changes; this is the advantage of two-dimensional material hetero-interfaces over traditional insulating barrier layers such as MgO.
\end{abstract}

\keywords{Spintronics, First-principles, Graphene}
\maketitle

\section{\label{sec:intro} Introduction}
Spintronics is a promising technology for developing high-density and low-power data storage, such as hard disk heads and magnetic random access memories (MRAMs), magnetic tunnel junctions (MTJs) are considered essential components of these devices~\cite{bhatti2017spintronics,
hirohata2020review, endoh2020recent, naganuma2023spintronics}.
This study focuses on a binary ferromagnetic alloy, iron--palladium (FePd), as a material for spintronics applications. It has an $L1_0$-ordered crystal structure~\cite{naganuma2015electrical, naganuma2020perpendicular, naganuma2022unveiling, uemoto2022density, naganuma2023jpcc, zhang2018enhancement, mohri2001theoretical, klemmer1995magnetic, shima2004lattice, miyata1990ferromagnetic, iihama2014low, kawai2014gilbert, itabashi2013preparation}.
$L1_0$-FePd exhibits a large magnetic anisotropy energy of $K_u \approx 10^7~\text{erg}/\text{cm}^3$\;~\cite{klemmer1995magnetic, shima2004lattice, miyata1990ferromagnetic} and a low Gilbert damping constant of thin films $\alpha \approx 10^{-2}$ \; \cite{iihama2014low, kawai2014gilbert}, which make it suitable for MRAM applications.
The metal oxide insulator MgO~\cite{hallal2013anatomy, lu2021comparison, robertson2023comparing} has been considered as the barrier layer of the FePd-based MTJ~\cite{itabashi2013preparation, parkin2004giant, naganuma2015electrical}.
However, a nearly $10~\%$ lattice constant mismatch can cause the system to face difficulties in obtaining a smooth interface and maintaining the high magnetoresistance (MR) ratio performance; as substitutes of MgO, two-dimensional (2D) materials, such as graphene and h-BN, are promising for use as barrier layer of MTJs~\cite{lu2021comparison, robertson2023comparing, lu2021ab};
such application of 2D materials in spintronics has been gaining interest in recent~\cite{yazyev2009magnetoresistive,luan2015tunneling,iqbal2018recent,hashmi2020graphene,piquemal2020spin,ahn20202d,Avsar2020review}.

Recently, we have studied the hetero-interface between the (001) surface of $L1_0$-ordered FePd and graphene (Gr), called FePd(001)/Gr.
This interface has been experimentally fabricated by the chemical vapor deposition technique~\cite{naganuma2020perpendicular, naganuma2022unveiling}.
Since graphene has a hexagonal honeycomb lattice and FePd has a square lattice, the FePd(001)/Gr interface exhibits a lattice symmetry mismatch, which is different from well-understood symmetry-matched metal--graphene interfaces, such as Ni(111)/Gr~\cite{kozlov2012bonding}.
In previous work, we performed first-principles calculations to predict the atomic structure, bonding mechanism, and electronic and magnetic properties of lattice symmetry-mismatched FePd/Gr interfaces~\cite{uemoto2022density, naganuma2023jpcc}.

In this study, we focus on a heterojunction composed of FePd electrodes and multilayer graphene, denoted as FePd/$m$-Gr/FePd, where $m$ signifies the number of graphene layers; we consider the case of monolayer ($m=1$), a bilayer ($m=2$), and a trilayer ($m=3$).
We perform first-principles predictions of spin-dependent transport properties for the proposed heterojunction and analyze spin-resolved conductance for various magnetic configurations.
The obtained MR ratio, in the range of $150$--$200~\%$, is not inferior to those of previously studied lattice symmetry-matched metal/graphene heterojunctions:
e.g., M/Gr/M (where M stands for Fe, Co, Ni, or Cr) exhibits $\mathrm{MR}=17~\% \sim 108~\%$
\cite{yazyev2009magnetoresistive,luan2015tunneling}.

In addition, we explore the effect of varying the number of graphene layers on the transport properties.
Furthermore, we examine the effect of lateral displacement (``sliding'') at the interface and find that the spin transport properties remain robust despite these modulations.
Moreover, since the junction incorporates lattice symmetry mismatch at the interfaces; understanding such systems may pave the way to new spintronics materials.
We further perform micromagnetic simulations of the FePd/1-Gr/FePd junction structure to investigate its spin-transfer torques (STT) switching characteristics.

The remainder of this paper is organized as follows: in Sec.~\ref{sec:method}, we present the interface structure models and the computational conditions.
In Sec.~\ref{sec:result}, we report the calculated spin-resolved transport properties and analysis of performance as MR .
Finally, in Sec.~\ref{sec:summary}, we present the summary.

\section{Method}
\label{sec:method}

\begin{figure}[htbp]
 \centering
 \includegraphics[width=0.45\textwidth]{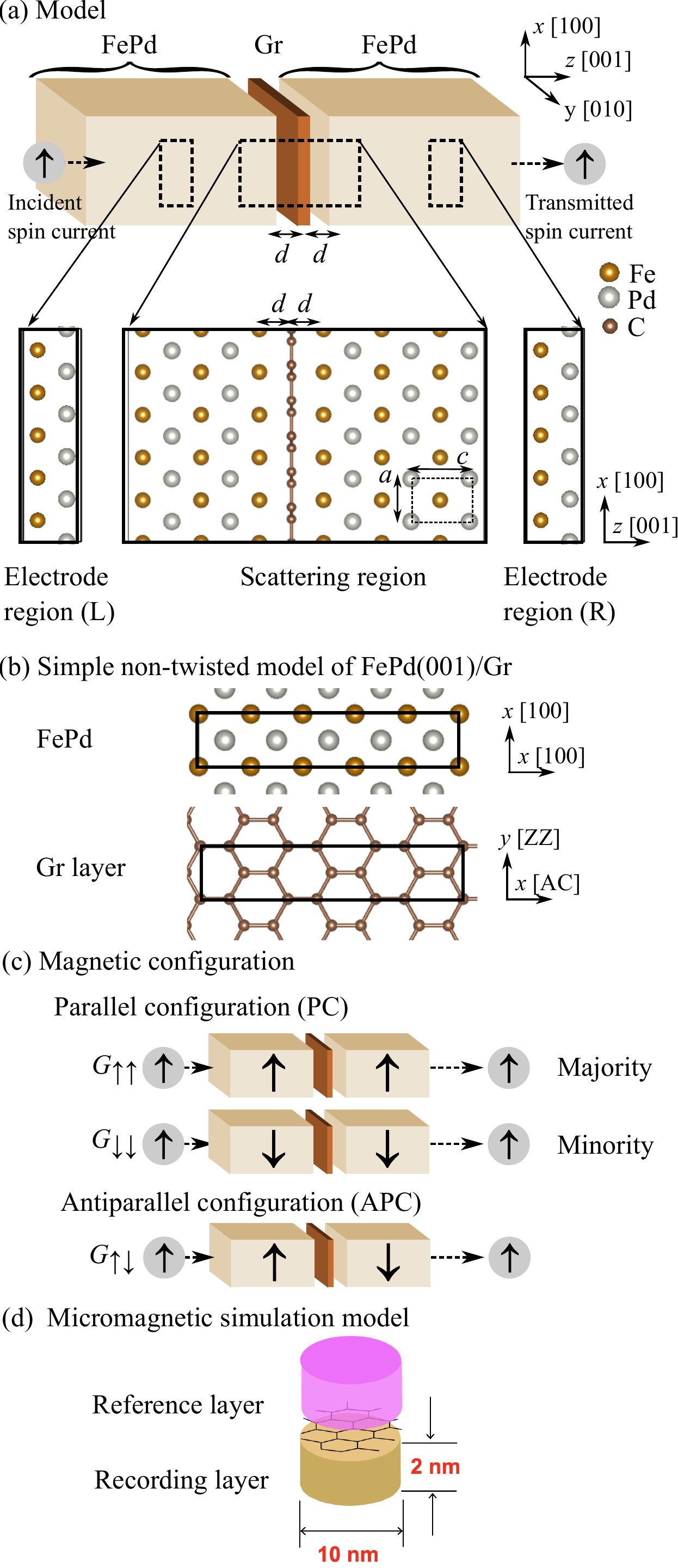}
 \caption{
 Schematic illustration of the spin transport calculation for the FePd/$1$-Gr/FePd heterojunction.
 (a) Computational model of the scattering region and electrode regions.
 (b) Cross-sectional view of the FePd(001)/Gr interface model (simple non-twisted model proposed in Ref.~\cite{uemoto2022density}).
 (c) Magnetic configurations and the spin-resolved conductance components $G_{\uparrow\uparrow}$, $G_{\downarrow\downarrow}$, and $G_{\uparrow\downarrow}$.
 (d) The micromagnetic simulation model of the junction consists of a FePd (recording) layer with a diameter of $10$~nm and a thickness of $2$~nm.
 The thick black-line boxes in (a) and (b) represent the supercells used in our calculations.
 }
 \label{fig:model}
\end{figure}

In this study, we consider first-principles calculations on those of FePd/$m$-Gr/FePd ($m = 1$--$3$) heterojunctions.
Figure~\ref{fig:model} shows a schematic illustration of monolayer graphene (FePd/1-Gr/FePd).
The computational model consists of a scattering region including the FePd(001)/Gr interfaces and electrode regions of bulk FePd, and the left and right electrodes surround the scattering region.
The incident spin-polarized current from the left electrode is transported to the right electrode.
In the scattering region, the monolayer graphene is sandwiched between three Fe atomic layers and two Pd layers on the left side and three Fe layers and three Pd layers on the right side.
We assume that the graphene layer is directly facing the Fe atomic layer.
Although numerous atomic scale structures can be considered, our Fe-top model agrees well with the results of previous experimental (by scanning transmission electron microscopy) and theoretical (by first-principles) works~\cite{naganuma2020perpendicular, uemoto2022density}.
For the atomic configuration of C atoms on the Fe layer, we employ the simple non-twisted model proposed in our previous work~\cite{uemoto2022density}.
The Fe--Gr interlayer distance is assumed to be $d \approx 2~\text{\AA}$.
As seen in Fig.~\ref{fig:model}(b), the armchair (AC) and zigzag (ZZ) axes of graphene are parallel to the $[100]$ and $[010]$ axes of FePd, respectively.
The lattice constant of FePd is assumed to be $a = 2.67~\text{\AA}$ and $c=3.70~\text{\AA}$ based on the optimization under equilibrium conditions~\cite{uemoto2022density}.
We consider a few different magnetic configurations in which the magnetization of the left and right electrodes is polarized in the parallel (parallel configuration: PC) or antiparallel (antiparallel configuration: APC) direction [see Fig.~\ref{fig:model}(c)].
We denote the majority and minority spin-resolved conductances in the PC as $G_{\uparrow\uparrow}$ and $G_{\downarrow\downarrow}$, respectively.
The conductance in APC is represented as $G_{\uparrow\downarrow}$.
The MR ratio can be expressed as ~\cite{hashmi2020graphene}
\begin{align}
 \text{MR}
 =& \frac{(G_{\uparrow\uparrow}+G_{\downarrow\downarrow})/2-G_{\uparrow\downarrow}}{G_{\uparrow\downarrow}}
 \times 100~\%
 \;.
 \label{eq:mr}
\end{align}
We use the RSPACE code for computation, which implements real-space first-principles electronic state and transport calculations based on density functional theory (DFT)~\cite{ono1999timesaving, ono2005real, tsukamoto2013tuning}.
We employ the local spin density approximation with the Vosko--Wilk--Nusair exchange-correlation functional~\cite{vosko1980accurate} and Troullier--Martins norm-conserving pseudopotentials~\cite{troullier1991efficient, NCPP}.
To solve the Kohn--Sham equation by the real-space approach, we use the finite-difference method with uniform and orthogonal mesh grids~\cite{chelikowsky1994finite, chelikowsky1994higher}.
For the electrode region, we use a $48.4~\text{\AA} \times 9.7~\text{\AA} \times 13.2~\text{\AA}$ rectangular supercell containing 10 atoms; the scattering region is $48.4~\text{\AA} \times 9.7~\text{\AA} \times 80.7~\text{\AA}$ with 67 atoms [see Fig.~\ref{fig:model}(a)].
The grid spacing is approximately $0.6~\text{\AA} \times 0.6~\text{\AA} \times 0.7 ~\text{\AA}$, which is dense enough for accurately representing the transition metal properties.
 Incidentally, the simulation of transport properties requires the ground state DFT results of the scattering region itself. To obtain these calculations, the supercell of the scattering region must satisfy periodic boundary conditions, which results in an asymmetric structure with a Fe layer at the left end and a Pd layer in the neighboring right end, as depicted in Fig.~\ref{fig:model}(a).

In addition, we also perform micromagnetic simulations of the FePd/Gr/FePd junction structure to predict the STT-induced magnetization reversal.
The simulation model, illustrated in Fig.~\ref{fig:model}(d), has a junction diameter of $10$~nm and the thickness of the recording layer $t_\mathrm{F} = 2$~nm.
To investigate the high-speed temporal changes in magnetization by the applied electric pulse, we numerically computed the following Landau-Lifshitz-Gilbert (LLG) equation:
\begin{align}
\frac{\partial \vec{M}_\text{F}}{\partial t}
=&
- \gamma \left[
 \vec{M}_\mathrm{F}
 \times
 \vec{H}_\mathrm{eff}
\right]
+
\frac{\alpha}{{M}_\mathrm{S}} \left[
 \vec{M}_\mathrm{F}
 \times
 \frac{\partial \vec{M}_\mathrm{F}}{\partial t}
\right]
+
\frac{\hbar}{2e} J_\mathrm{e} g(\theta)
\left[
\vec{M}_\mathrm{F} \times (\vec{M}_\mathrm{R} \times \vec{M}_\mathrm{F})
\right]
\;,
\label{eq:llg}
\end{align}
where $\vec{M}_\mathrm{F}$ and $\vec{M}_\mathrm{R}$ denote the magnetization vectors of the recording and reference layers, respectively; ${M}_\mathrm{S}$ represent the saturation magnetization for the recording layer. Furthermore, $\gamma$ is the gyromagnetic ratio, $\vec{H}_\mathrm{eff}$ is the effective magnetic field, $J_e$ is the current density, and $g(\theta)$ is the spin-transfer efficiency (the details are given in Ref.~\cite{naganuma2023spintronics}).
The computations are performed using the Fujitsu EXAMAG LLG simulator\cite{yoshida2022field}.

\clearpage
\section{Results}
\label{sec:result}

\begin{figure*}[htbp]
 \centering
 \includegraphics[width=0.75\textwidth]{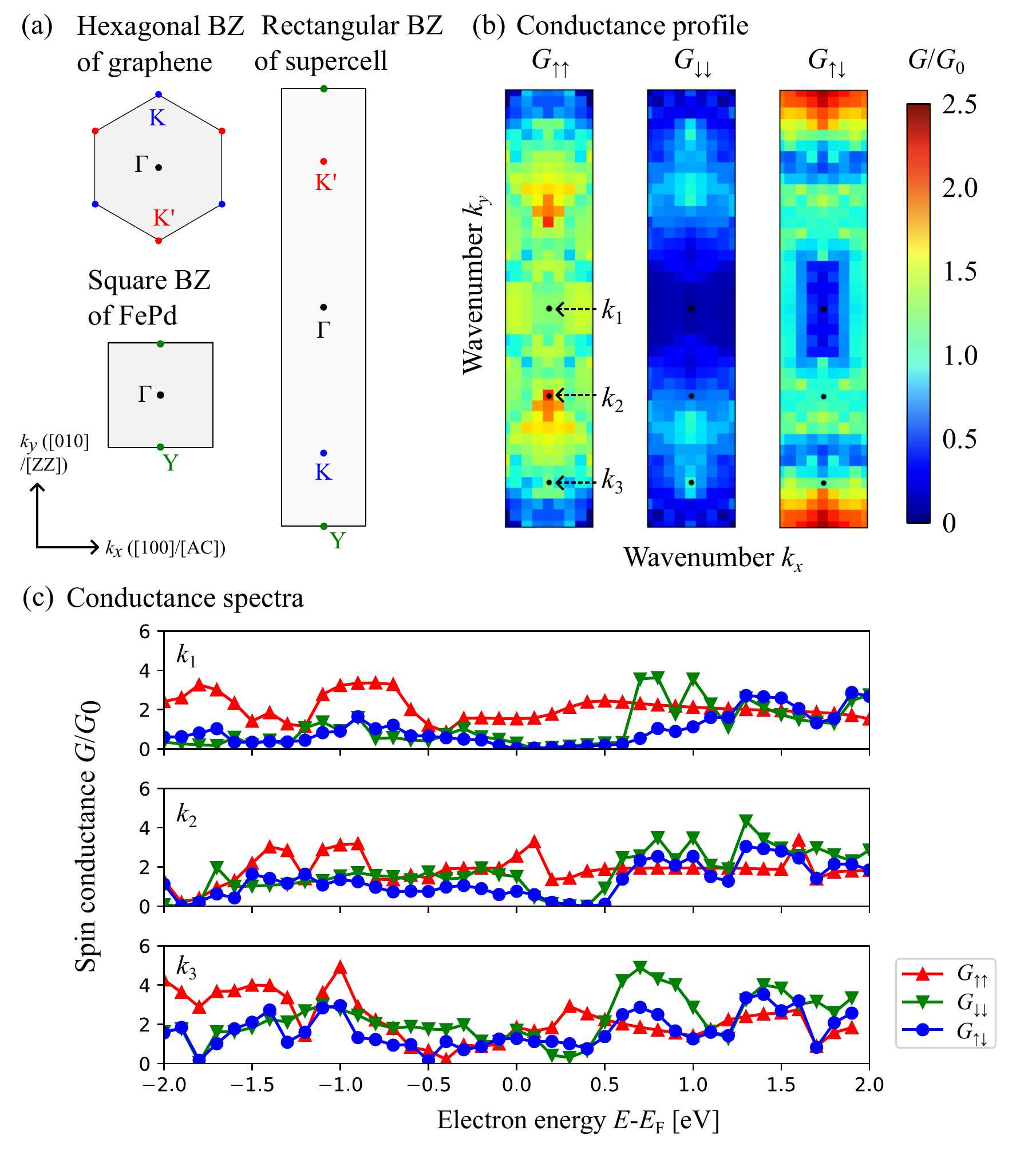}
 \caption{
 Spin-resolved conductance, $k$-space profile, and spectra of FePd/1-Gr/FePd heterojunction.
 (a) BZ for graphene (hexagonal BZ), bulk FePd (square BZ), and slab supercell (rectangular BZ).
 (b) $k$-space profiles of $G_{\uparrow\uparrow}$, $G_{\downarrow\downarrow}$, and $G_{\uparrow\downarrow}$ as a function of transverse wavevectors $k_x, k_y$ at the Fermi level $E=E_\mathrm{F}$.
 The labels $k_1$, $k_2$, and $k_3$ correspond to the $k$-points selected from the $\Gamma-K-Y$ segment.
 (c) Spin-resolved conductance spectra: $G_{\uparrow\uparrow}$, $G_{\downarrow\downarrow}$, and $G_{\uparrow\downarrow}$ as a function at $k_1$, $k_2$, and $k_3$.
 }
 \label{fig:profile}
\end{figure*}

First, for simplicity, we confine our discussion to the monolayer-graphene-based heterojunction (FePd/1-Gr/FePd) and compute spin-dependent transport properties.
This calculation provides the spin-resolved conductance $G_{\sigma\sigma'}(E; k_x, k_y)$, dependent on the electron energy $E$, transverse wave vectors $k_x, k_y$ and spin indices $\sigma, \sigma'$.
In Fig.~\ref{fig:profile}(a), we illustrate the Brillouin zone (BZ) of graphene (hexagonal BZ) and that of bulk FePd (square BZ) for well-known primitive cells.
Additionally, we illustrate the BZ for the supercell (rectangular BZ).
Conventionally, the $\Gamma$ and $K$($K'$) points are placed at the center and vertices of the hexagonal BZ, respectively.
In the rectangular BZ, the $\Gamma$ and $K$($K'$) points are mapped onto a segment along the $k_y$-axis.
In Fig.~\ref{fig:profile}(b), we plot the $k$-space profiles of the spin-resolved conductance: $G_{\uparrow\uparrow}$, $G_{\downarrow\downarrow}$, and $G_{\uparrow\downarrow}$ components as functions of $k_x$ and $k_y$.
The averaging of energy $E$ is performed around the Fermi level ($E_\mathrm{F}$) within the range $E_\mathrm{F}-\Delta{E} \sim E_\mathrm{F}+\Delta{E}$ as below:
\begin{align}
 G_{\sigma \sigma'}(k_x, k_y)
 =&
 \frac{1}{2 \Delta{E}} \int_{E_F-\Delta{E}}^{E_F+\Delta{E}}
 G_{\sigma \sigma'}(E; k_x, k_y) \; \mathrm{d}E
 \;,
 \label{eq:avg_e}
\end{align}
where $\Delta{E}=0.01~\text{eV}$ in Fig.~\ref{fig:profile}(b).
Additionally, the normalization of plots is performed using the conductance quantum constant $G_0 \equiv 2e^2/h \approx 7.748 \times 10^{-5}$~$\Omega^{-1}$.

The conductance is distributed over a broad area of the BZ (region spanning between the $\Gamma$ and $K$ points). $G_{\uparrow\uparrow}$ is larger than the minority spin ($G_{\downarrow\downarrow}$), which reflects the density of states (DoS) at $E_\mathrm{F}$.
For a more detailed understanding, we consider the conductance at selected $k$ points: $k_1 \equiv (0, 0)$, $k_2 \equiv (0, \pi/5L_y)$, and $k_3 \equiv (0, 3\pi/5L_y)$, where $L_x$ and $L_y$ represent the size of the supercell.
Figure~\ref{fig:profile}(c) presents the spin-resolved conductivity spectra: $G_{\uparrow\uparrow}$, $G_{\downarrow\downarrow}$, and $G_{\uparrow\downarrow}$ as functions of energy $E$ at $k_1$, $k_2$, and $k_3$.

We also consider the conductance averaged over the BZ, expressed as
\begin{align}
G_{\sigma \sigma'} =& \frac{1}{\Omega_\mathrm{BZ}}
\iint_\mathrm{BZ}
G_{\sigma \sigma'}(k_x, k_y)
\;
\mathrm{d} k_x
\mathrm{d} k_y
\;,
\label{eq:avg_g}
\end{align}
where $\Omega_\mathrm{BZ}$ is the area of the BZ.
The $k$-averaged values are $G_{\uparrow\uparrow} \approx 1.4G_0$, $G_{\downarrow\downarrow} \approx 1.3G_0$, and $G_{\uparrow\downarrow} \approx 0.5G_0$.
The MR ratio~(\ref{eq:mr}) is calculated to be $\approx 150~\%$. This value does not significantly degrade compared with those previously reported for other metal/graphene interfaces: e.g., $\mathrm{MR}=61~\%$ has been reported for Fe/Gr/Fe, $\mathrm{MR}=60~\%$ for fcc Co/Gr/Co, $\mathrm{MR}=86~\%$ for hcp Co/Gr/Co, and $\mathrm{MR}=17~\%$ for Ni/Gr/Ni in Ref.~\cite{yazyev2009magnetoresistive}. For Ni/Gr/Ni, $\mathrm{MR}=60~\%$ has been reported~\cite{luan2015tunneling}, as well as $\mathrm{MR}=108~\%$ for Cr/Gr/Cr.

\begin{figure*}[htbp]
\centering
\includegraphics[width=1.0\textwidth]{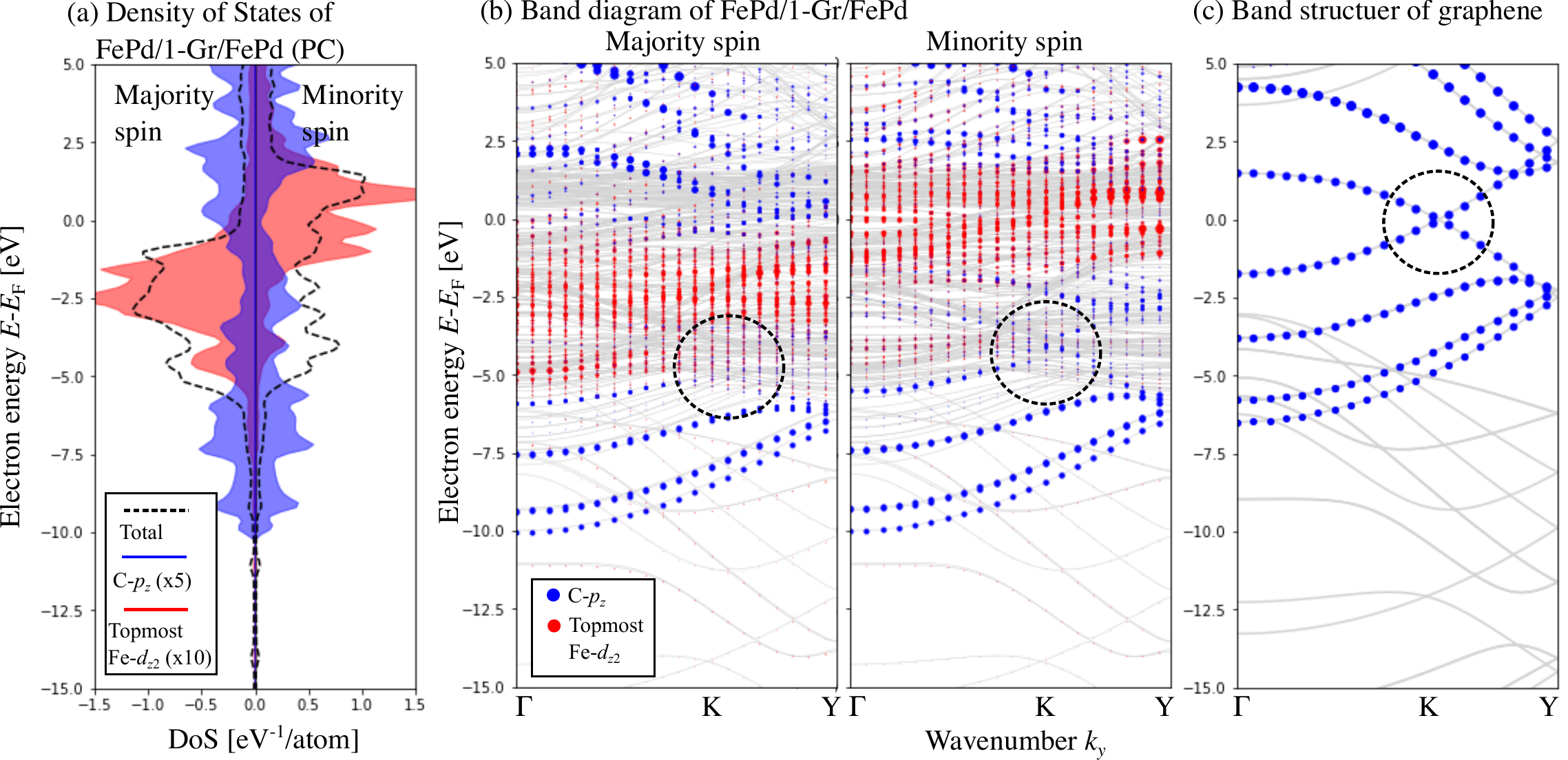}
\caption{
 DoS and electronic band structure of FePd/1-Gr/FePd (PC) interface:
 (a) total and partial DoS curves.
 The blue and red areas represent the partial DoSs projected onto the $p_z$ orbital of C atoms in the graphene layer (C-$p_z$) and the $d_{z^2}$ orbital of Fe atoms in the topmost Fe layer (topmost Fe-$d_{z^2}$).
 (b) Band structure of FePd/1-Gr/FePd.
 The horizontal axis of the band diagrams represents
 the line segment through $\Gamma - K - Y$ in the rectangular BZ in Fig.~\ref{fig:profile}(a).
 (c) Band structure of isolated graphene.
 The red- and blue-filled circles superposed on the bands represent the weights of C-$p_z$ and the topmost Fe-$d_{z^2}$ contributions, respectively.
}
\label{fig:band}
\end{figure*}

Figure~\ref{fig:band} presents the density of states (DoS) and band structures of the FePd/1-Gr/FePd interface (with the PC magnetization) as well as those of isolated monolayer graphene.
These calculations are performed using the Vienna Ab-initio Software Package (VASP) code.
Detailed information about the computational conditions is provided in Supplemental Material.
 In Fig.~\ref{fig:band}(a), the black curve represents the total DoS; the red and blue areas illustrate the partial DoSs projected onto the $p_z$ orbital of C atoms in the graphene layer (C-$p_z$) and onto the $d_{z^2}$ orbital of Fe atoms in the topmost Fe layer (topmost Fe-$d_{z^2}$).

As seen in the DoS, the $d_{z^2}$ band shows well-defined exchange splitting.
The $d_{z^2}$ band for the majority spin is located beneath the Fermi level and is largely occupied.
In the conductance spectra of Fig.~\ref{fig:profile}(c), the majority spin component ($G_{\uparrow\uparrow}$) increases for the lower energy region ($E<E_\mathrm{F}$), while the minority spin component ($G_{\downarrow\downarrow}$) increases for the higher energy region ($E>E_\mathrm{F}$).
This behavior reflects the spectral shape of the spin-resolved DoS, which suggests that the conduction in junctions including graphene layers is dominated by the tunneling conduction between the FePd electrodes.
Figure~\ref{fig:band}(b) illustrates the band structure along the line segment through the $\Gamma-K-Y$ of the rectangular BZ in Fig.~\ref{fig:profile}(a).
The red- and blue-filled circles superposed on the bands represent the weights of C-$p_z$ and the topmost Fe-$d_{z^2}$ contributions, respectively.
For convenience, we also display the band structure of isolated graphene in Fig.~\ref{fig:band}(c).
Charge transfer from the FePd surface to the graphene results in a shift of the $p_z$ band toward lower energy levels.
In graphene, the Dirac cone at the $K$ point plays a crucial role in electron transport.
However, in the coupled system, the Dirac cone is strongly hybridized with the $d_{z^2}$ band, and its band edge is positioned deeply below the Fermi level (approximately at $-5~\mathrm{eV}$), as highlighted by the dotted circles in Figs.~\ref{fig:band}(b) and \ref{fig:band}(c).

The C-$p_z$-based, $\pi$-band, which is on the high energy conduction band of the Dirac cone, plays a crucial role in the transport around the Fermi level.

\begin{figure}[htbp]
 \centering
 \includegraphics[width=0.4\textwidth]{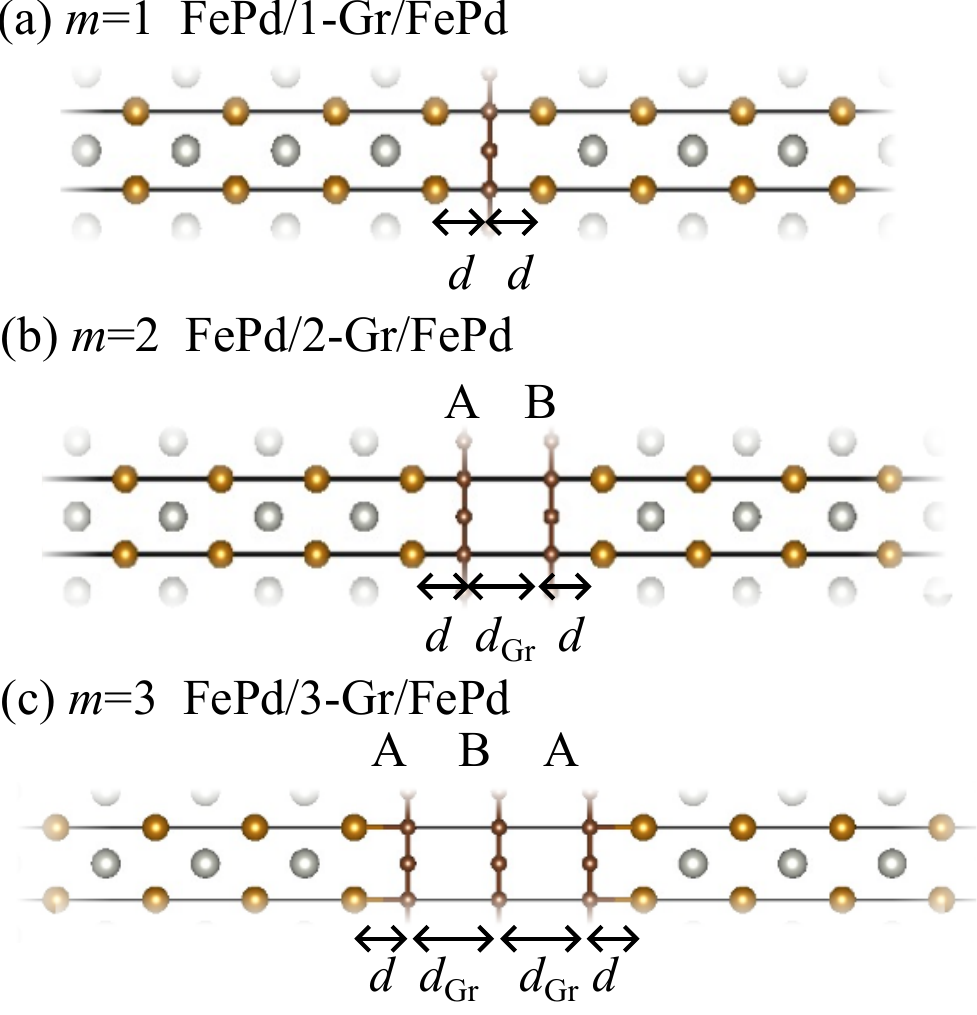}
 \caption{Illustration of FePd/multilayer Gr/FePd (FePd/$m$-Gr/FePd) heterojunction with $m=1 \text{--} 3$.}
 \label{fig:multilayer}
\end{figure}

\begin{figure}[htbp]
 \centering
 \includegraphics[width=0.45\textwidth]{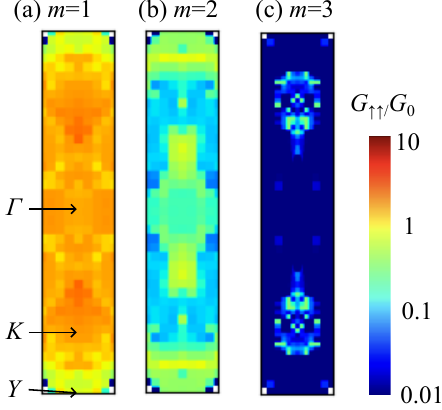}
 \caption{
 $k$-space profiles of spin-resolved conductance $G_{\uparrow\uparrow}$ for multilayer graphene-based FePd/$m$-Gr/FePd heterojunction:
 (a) monolayer graphene ($m=1$),
 (b) bilayer graphene ($m=2$), and
 (c) trilayer graphene ($m=3$).
 }
 \label{fig:profile_multi}
\end{figure}

\begin{table}[h]
\caption{\label{tbl:g}
 Spin-resolved conductance of FePd/$m$-Gr/FePd heterojunction with the monolayer ($m=1$), bilayer ($m=2$), and trilayer ($m=3$) graphene shown in Fig.~\ref{fig:multilayer}.
 $G_{\sigma\sigma'}$ is averaged over $k$ and $E$ using Eqs.~(\ref{eq:avg_e}) and (\ref{eq:avg_g}), the parameter $\Delta{E}$ in the averaging is assumed to be $0.01$~eV (the values in parentheses are calculated by $\Delta{E}=0.03$~eV for comparison).
}
\begin{ruledtabular}
\begin{tabular}{lcccc}
\# of Gr layers($m$) &
Stacking order &
$G_{\uparrow\uparrow}$ &
$G_{\downarrow\downarrow}$ &
$G_{\uparrow\downarrow}$\\
\hline
1 (monolayer) & A & $1.4 G_0$ & $1.3 G_0$ & $0.5 G_0$ \\
 & & ($1.5 G_0$) & ($1.1 G_0$) & ($0.6 G_0$)\\
2 (bilayer) & AB & $0.3 G_0$ & $0.3 G_0$ & $0.1 G_0$\\
 & & ($0.3 G_0$) & ($0.3 G_0$) & ($0.1 G_0$)\\
3 (trilayer) & ABA & $\sim 2 \times 10^{-2} \; G_0$ & $\sim 3 \times 10^{-2} \; G_0$ & $\sim 1 \times 10^{-2} \; G_0$ \\
 & & ($\sim 3 \times 10^{-2} \; G_0 $) & ($\sim 2 \times 10^{-2} \; G_0$) & ($\sim 1 \times 10^{-2} \; G_0$) \\
\end{tabular}
\end{ruledtabular}
\end{table}

Incidentally, the spin- and $k$-space-resolved conductance profile usually reflects the physical mechanism underlying MR.
As seen in Fig.~\ref{fig:profile}(c), there are hotspots in the majority spin conductance around $k_3$ point at $\Gamma$-K-Y segment.
Previous pioneering studies for the large TMR effects in Fe/MgO/Fe(001) junctions have reported a single peak in the majority spin at $\Gamma$ to \cite{butler2001spin, mathon2001theory}, which reflects the existence of coherent tunneling processes.
Another mechanism to achieve large MR effects is the electron tunneling of interfacial electronic states; this is observed as ring-shaped conductance distributions around $\Gamma$ point \cite{masuda2021interfacial}.
Our system's profile can be considered to as the former type.
Additionaly, as seen in Fig.~\ref{fig:band}, at the FePd/$1-$Gr interface, the majority spin band dispersion of graphene intersects $E_F$ near the K point; this is level is originating to the conduction band of graphene shifted to lower energy by the charge transfer.
Such a conductance mechanism can be interpreted as a coherent tunneling mechanism rather than interfacial tunneling.

Next, we make on to multilayer graphene-based heterojunctions: FePd/$m$-Gr/FePd with monolayer ($m=1$), bilayer ($m=2$), and trilayer ($m=3$) graphene [see Fig.~\ref{fig:multilayer}(a)--\ref{fig:multilayer}(c)].
In our previous work~\cite{uemoto2022density}, the ABA stacking order of graphene was found to be the most stable configuration on FePd, where the interlayer distance is $d_\mathrm{Gr} \approx 3.3~\text{\AA}$.
The calculated spin-resolved conductance is shown in Table.~\ref{tbl:g}.
For the bilayer graphene ($m=2$), it can reach up to approximately $200~\%$.
However, as $m$ increases, the conductance decreases markedly; such exponential decay is also reported in Ni/$m$-Gr/Ni~\cite{karpan2008theoretical}.
In the case of trilayer graphene ($m=3$), because of the small conductance values (on the order of $10^{-2} \; G_0$), the accurate estimation of MR excluding numerical errors is not straightforward; we expect that the MR ratio on the order of $10^2$ is likely maintained.

\begin{figure}[htbp]
 \centering
 \includegraphics[width=0.49\textwidth]{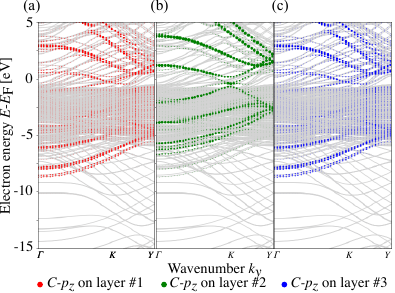}
 \caption{
 Majority spin band structure of FePd/3-Gr/FePd (trilayer graphene).
 The red, green, and blue dots indicate the weight of the C-$p_z$ orbital contribution in the first, second, and third graphene layers, respectively. Each layer corresponds to the left, middle, and right graphene in Fig.~\ref{fig:multilayer}(c), respectively.
 }
 \label{fig:band_layer}
\end{figure}

In Fig.~\ref{fig:profile_multi}, we present the $k$-space profiles of $G_{\uparrow\uparrow}$ for $m=1 \text{--} 3$.
For $m=1$, the conductance is dispersed over a large region of the BZ.
This behavior indicates the dominance of tunneling conduction between the metallic electrodes.
For $m=3$, the conductance appears in a very restricted area near the $K$ point.
In addition, the band structure of FePd/3-Gr/FePd is depicted in Fig.~\ref{fig:band_layer}.
The red, green, and blue dots represent the weights of the C-$p_z$ orbital contributions in the first, second, and third graphene layer, which correspond to the left, middle, and right graphene layers in Fig.~\ref{fig:multilayer}(c), respectively.
The first and third layers, which are in contact with the Fe surface, indicate the hybridization and charge transfer phenomena seen in the monolayer (m=1) case.
The second layer, which neighbors to other graphene layers via van der Waals force, exhibits a band structure that resembles that of isolated graphene, as seen in Fig.~\ref{fig:band}(c)].
Near the Fermi energy, there are conduction band minima of the Dirac cone at the K point.
When $m=3$, it is considered that interlayer conduction within a limited $k$-space in multilayer graphene dominates the electron transport.

\begin{figure}[htbp]
 \centering
 \includegraphics[width=0.45\textwidth]{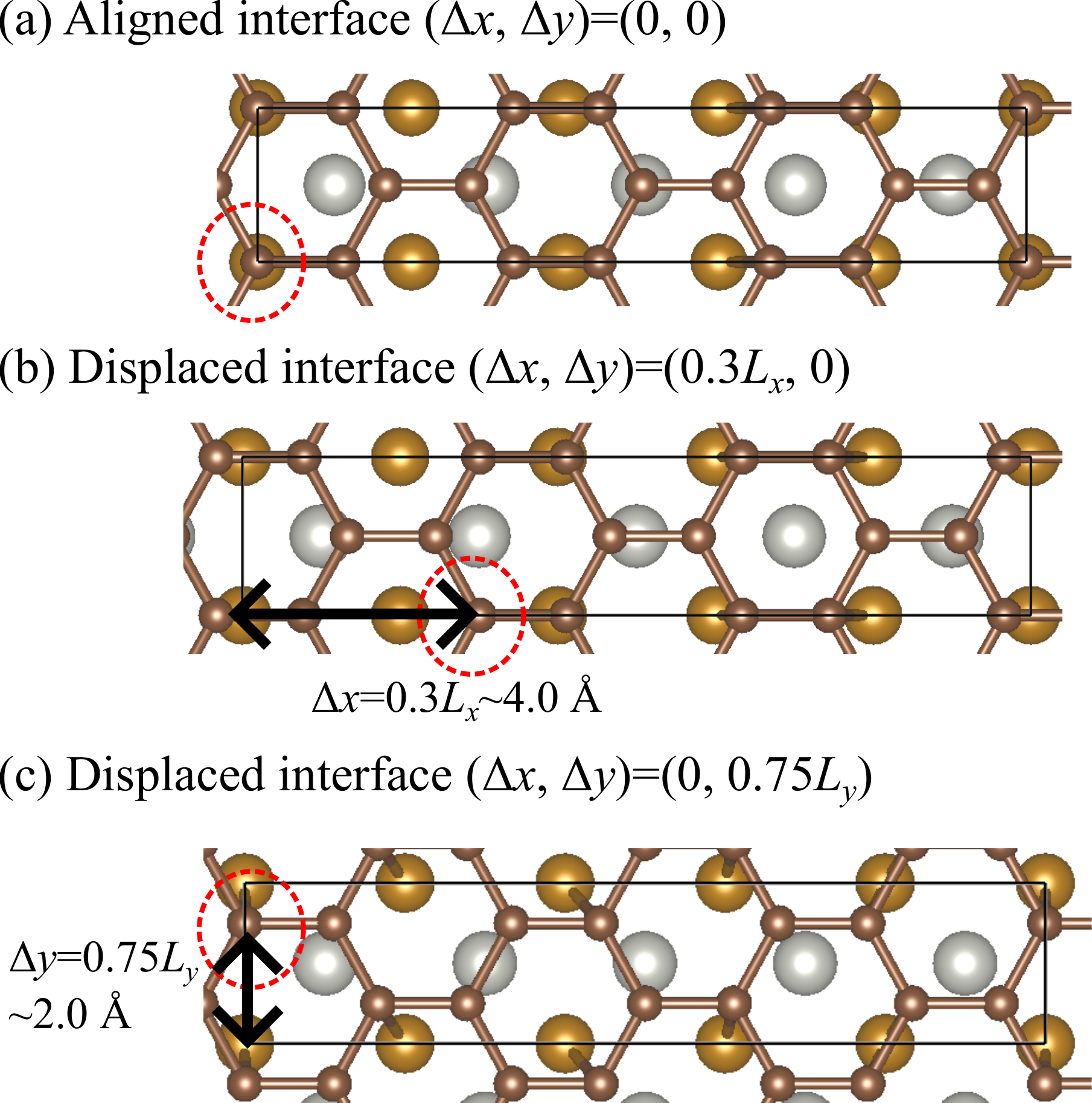}
 \caption{
 Schematic illustration of the FePd/Gr/FePd interface model with lateral slides (displacements).
 (a) A single C atom (marked by a dotted red circle) is positioned on a top of Fe atom (aligned structure).
 (b) $x$-axis displacement of $\Delta{x} = 0.3L_x \approx 4.0~\text{\AA}$
 (c) $y$-axis displacement of $\Delta{y} = 0.75a_y \approx 2.0~\text{\AA}$
 (The displacement vector of the marked C atom from the aligned interface structure is expressed as
 $(\Delta{x}, \Delta{y})$.)
 }
 \label{fig:displacement}
\end{figure}

\begin{table}[h]
\caption{\label{tbl:displacement}
Spin-resolved conductance $G_{\uparrow\uparrow}$, $G_{\downarrow\downarrow}$, and $G_{\uparrow\downarrow}$ of the monolayer graphene-based FePd/1-Gr/FePd heterojunction with lateral displacement $(\Delta{x}, \Delta{y})$ of Gr.
The detailed structure of the models (a)-(c) is shown in Fig.~\ref{fig:displacement}.}
\begin{ruledtabular}
\begin{tabular}{cllllll}
Model & $\Delta{x}$ & $\Delta{y}$ & $G_{\uparrow\uparrow}$ &
$G_{\downarrow\downarrow}$ &
$G_{\uparrow\downarrow}$ \\
\hline
(a) & 0 & 0 & $1.4 G_0$ & $1.3 G_0$ & $0.5 G_0$ \\
(b) & $0.3L_x (\sim 4.0~\text{\AA})$ & $0$ & $1.4 G_0$ & $1.4 G_0$ & $0.5 G_0$ \\
(c) & 0 & $0.75L_y (\sim 2.0~\text{\AA})$ & $1.4 G_0$ & $1.4 G_0$ & $0.7 G_0$
\end{tabular}
\end{ruledtabular}
\end{table}
\begin{figure}[htbp]
 \centering
 \includegraphics[width=0.4\textwidth]{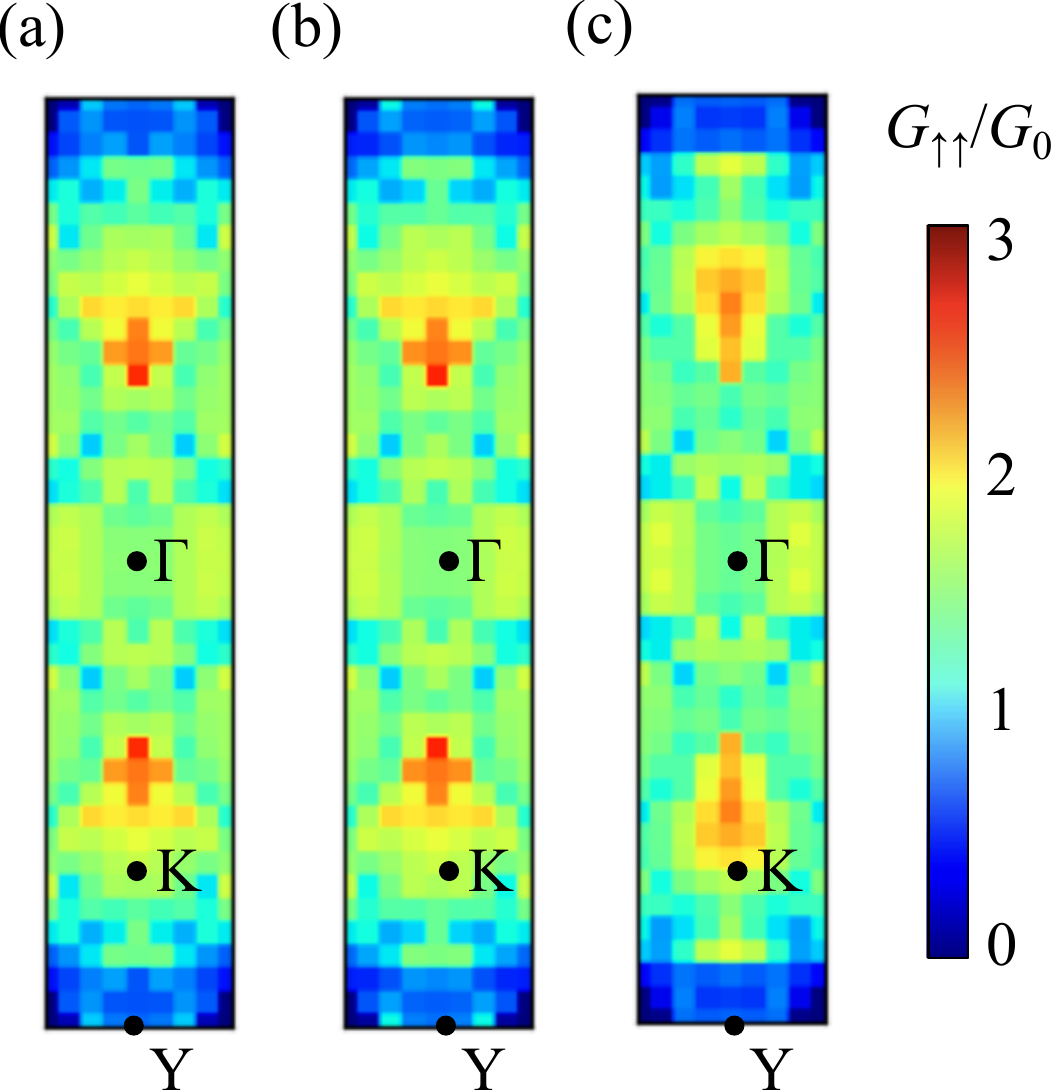}
 \caption{
 $k$-space profile of spin-resolved conductivity $G_{\uparrow\uparrow}(k_x, k_y)$
 for the FePd/$1$-Gr/FePd junction with different displacement (a)-(c), which corresponds to the aligned and displaced interfaces in Fig.~\ref{fig:displacement} (a)-(c).
}
 \label{fig:profile_slide}
\end{figure}

Next, we consider the stability of the spin-transport properties under the interface's displacement.
In our previous work~\cite{uemoto2022density}, we investigated the stability of the bonding states in FePd/Gr against lateral displacements (sliding) of the Gr layers, as illustrated in Fig.~\ref{fig:displacement}.
The obtained large binding energy, on the order of $10^2$~meV/atom, and the relatively small barrier for displacement (less than 1 meV/atom), suggest that sliding can easily occur within the FePd/Gr interface.
In this study, we extend our analysis to examine the effect of sliding on spin transport properties.
Here, we consider three different types of sliding, characterized by the lateral displacement of $(\Delta{x}, \Delta{y})$ [see Figs.~\ref{fig:displacement}(a)--(c)].
Figure~\ref{fig:profile_slide} shows the $k$-space profile of the $G_{\uparrow\uparrow}$ component.
The sliding of the interface has a negligible effect on the distribution of conductance in the $k$-space.
The averaged conductance values determined using Eqs.~(\ref{eq:avg_e}) and (\ref{eq:avg_g}) are shown in Table.~\ref{tbl:displacement}.
There are only small qualitative differences, and the MR ratio is maintained.
This behavior can be understood as originating from the lattice symmetry mismatch between FePd and Gr.

In symmetry-matched metal/Gr interfaces, such as Ni(111)/Gr, C atoms are positioned at specific adsorption sites on the metal surface, resulting in a constant relative position between metal and C atoms.
In such systems, an obvious site dependence of electronic states and bonding mechanisms has been reported~\cite{kozlov2012bonding}, and we can consider that sliding at the interface could induce significant changes in electronic transport properties~\cite{robertson2023comparing}.
On the other hand, in lattice-mismatched systems (FePd/Gr), various relative atomic positions exist on the interface and the averaged interatomic interactions remain unchanged upon displacement.
Such behavior in the electronic and magnetic states was discussed in our previous work~\cite{uemoto2022density}; the same argument is also extended to analyzing transport properties.
This robustness of the spin transport properties can be expected to be advantageous.

 Incidentally, the above calculations assume a ground state (at zero temperature).
However, the spectra shown in Fig.~\ref{fig:profile}(c) exhibit that the behavior remains nearly constant within the range of $E_\text{F} \pm 0.1~\text{eV}$. ; this suggests that minor thermal excitations, such as room temperature ($k_\text{B} T_\text{Room} \approx 0.025~\text{eV} \ll 0.1~\text{eV}$), does not significantly modify the microscopic scale spin transport properties.

\begin{figure*}[htbp]
 \centering
 \includegraphics[width=0.95\textwidth]{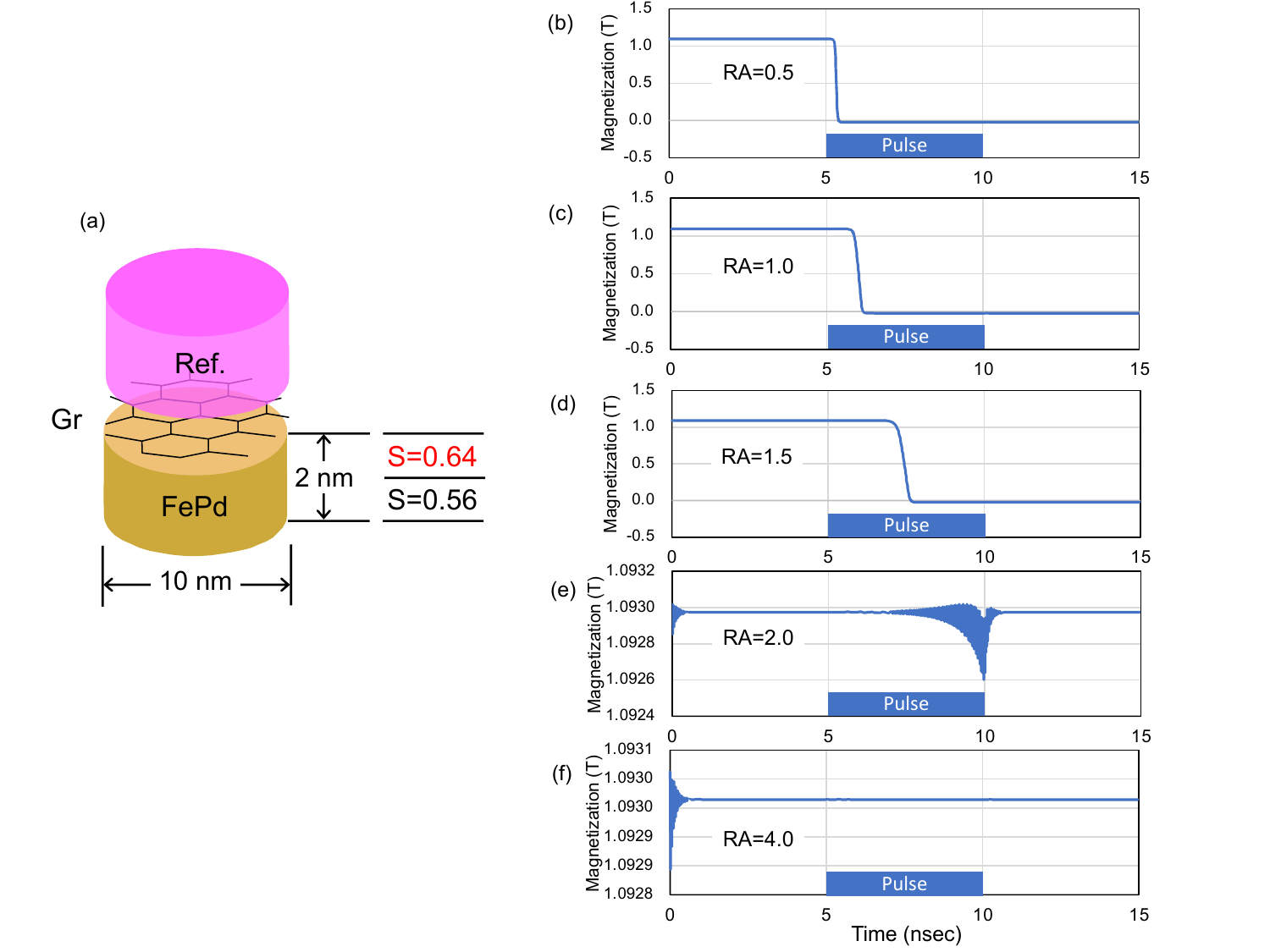}
 \caption{
 Micromagnetic simulation of the STT magnetization switching curve.
 }
 \label{fig:micromag}
\end{figure*}

We also perform the micromagnetic simulation of the high-speed STT magnetization switching behavior in the FePd/1-Gr/FePd junctions, which diameter of $10$~nm and a thickness of $t_\mathrm{F}=2$~nm [see also Fig.~\ref{fig:micromag}(a)]; they are designed such that their thermal stability ($\Delta$) can achieve $\Delta \ge 57$.
Figure~\ref{fig:micromag} represents the magnetization switching curve (change in perpendicular magnetization $M_z$ as a function of time) for various spin-polarization $S$ and resistance-area product ($RA$) values.
The spin polarization value is defined as
$S=[D_\uparrow(E_\mathrm{F}) -D_\downarrow(E_\mathrm{F})]/[D_\uparrow(E_\mathrm{F}) +D_\downarrow(E_\mathrm{F})]$, where $D_\uparrow(E_\mathrm{F})$ and $D_\downarrow(E_\mathrm{F})$ represent the majority and minority spin components of the DoS at the Fermi level, respectively; the first-principle calculation provides $S=0.63$ for the FePd at Gr interface and $S=0.59$ for the bulk FePd.
We assume interfacial spin polarization of $S = 0.63$ for 1~nm thick from the Gr interface, and the other 1~nm is set to $S = 0.59$ as bulk spin polarization [Fig.~\ref{fig:micromag}(b)-(f)].
The experimentally reported value of $RA$ for the graphene tunnel barrier at room temperature is $RA = 1.5~\Omega\cdot\mu\mathrm{m}^2$ in Ref.~\cite{li2014magnetic}.
 The parameters obtained from first-principles calculations assume the ground state, but as mentioned earlier, they can be justified even under weak thermal excitations at room temperature.
For convenience, we consider several possible values for $RA$: 0.5, 1.0, 1.5, 2.0, and 4.0.
We set the saturation magnetization $M_\mathrm{S}$ in Eq.~(\ref{eq:llg}) to $1.2$~T.
The applied pulse voltage is $1$~V, and the pulse duration is $5$~nsec.
For small $RA$, the magnetization reversal time of the FePd layer decreases.
Here we consider the following relation:
\begin{align}
 V_\mathrm{C} =&
 \frac{2 e \alpha}{\hbar S}
 \; RA \;
 \mu_0
 M_\mathrm{S}
 t_\mathrm{F}
 H_k
 \;,
\end{align}
where
$V_\mathrm{C}$ is the critical write voltage and $H_k$ is the effective anisotropy field.
The decrease in RA led to an increase in the spin-polarized current at the same voltage, resulting in a reduction in $V_\mathrm{C}$. By lowering $RA$, which is a characteristic of Gr, and increasing interfacial spin polarization, nanosecond high-speed switching can be realized at a relatively low pulse voltage of 1~V.

\clearpage
\section{Summary}
\label{sec:summary}
We presented first-principles predictions of spin-dependent transport properties for a heterojunction consisting of ferromagnetic alloy FePd and multilayer graphene (FePd/$m$-Gr/FePd).
 This junction structure is currently in the experimental design phase; the observation has not been obtained yet.
We considered the interface structure model based on the simple non-twisted model proposed in our previous theoretical work~\cite{uemoto2022density}.
The MR ratio of the FePd/$m$-Gr/FePd junction was estimated to be approximately $150 \text{--} 200~\%$.
Despite the lattice symmetry mismatch at our interface between FePd (square lattice) and graphene (hexagonal honeycomb lattice), the MR ratio of the FePd/$m$-Gr/FePd system did not significantly fall below that of previously studied lattice symmetry-matched metal/graphene heterojunctions:
M/Gr/M (where M stands for Fe, Co, Ni, or Cr) exhibited $\mathrm{MR}=17~\% \text{--} 108~\%$~\cite{yazyev2009magnetoresistive, luan2015tunneling}.
Thus, we can suggest that the absence of lattice symmetry does not inherently degrade the MR performance.

In addition, an increase in the number of graphene layers $m$ resulted in a reduction in conductance and a shift in the dominant transport mechanism from electrode's tunnel conductance to graphene's interlayer transition around the $K(K')$ point; therefore, a smaller $m$ value is desired for the MR application.
Moreover, in interfaces with lattice symmetry mismatches, the averaging of interatomic interactions led to the loss of site dependence in the electronic structure~\cite{uemoto2022density}.
The transport properties persisted despite interface sliding (lateral displacement).
This robustness of the MR performance for such modulations is considered to be an advantage over ordinary barrier materials such as MgO.

Generally, the MR effects of the junction structures can be divided into two major categories: giant magnetoresistance (GMR) and tunneling magnetoresistance (TMR). The GMR is associated with metallic barriers and TMR with insulating barriers.
In the case of FePd/1-Gr/FePd, the majority-spin band dispersion in Fig.~\ref{fig:band} shows that the bands originating from the conduction band of graphene are shifted to lower energy and intersect at $E_\mathrm{F}$.
This configuration enables graphene to exhibit conductive properties, which suggests that the junction structures exhibit GMR-like responses.
However, our system's order of $MR \approx 10^2~\%$ which overcomes the typical values in GMR systems.
Additionary, similar Ni/Gr/Ni junctions have been categorized as GMR \cite{yazyev2009magnetoresistive}.

As number of layers $m$ increases, a bandgap-like electronic structure emerges near $E_\mathrm{F}$ [see Fig.~\ref{fig:band_layer}(b)].
This occurs due to the weak interactions between neighboring layers or with the metal surface, causing a perturbative split in the degenerate states at the Dirac point. This suggests the potential for TMR-like mechanisms in multi-layered structures.
Therefore, we consider that our FePd/$m$-Gr/FePd system does not fit into the typical GMR classification and exhibits characteristics that are intermediate between GMR and TMR.

Heterojunctions incorporating lattice symmetry mismatch have not been well studied yet.
We expect that gaining an understanding of such systems may pave the way to new spintronics materials.

\section*{SUPPLEMENTARY MATERIAL}
See the supplementary material for computational conditions for calculating the density of states and band structure (S1).

\begin{acknowledgments}
This study is partly supported by the Japan Society for the Promotion of Science (JSPS)
 Core-to-Core Program (No. JPJSCCA20230005), by Cooperative Research Project from
CSRN, and by the cross-appointment project (H.N, P.S., and J.R.) and QST-Tohoku University matching foundation.
In addition, this work is also partially financially supported by MEXT as part of the ``Program for Promoting Researches on the Supercomputer Fugaku'' (Quantum-Theory-Based Multiscale Simulations toward the Development of Next-Generation Energy-Saving Semiconductor Devices, JPMXP1020200205), JSPS KAKENHI (JP22H05463), JST CREST(JPMJCR22B4), Kurata Grants, and the Iwatani Naoji Foundation. The numerical calculations were carried out using the computer facilities of the Institute for
Solid State Physics at The University of Tokyo, the Center for Computational Sciences at the University of Tsukuba (Multidisciplinary Cooperative Research Program), and the supercomputer Fugaku provided by the RIKEN Center for Computational
Science (Project ID: hp210170, hp230175).
The visualization is performed by VESTA code~\cite{momma2008vesta}.
\end{acknowledgments}

\section*{AUTHOR DECLARATIONS}
\subsection*{Conflict of Interest}
The author has no conflicts to disclose.

\section*{DATA AVAILABILITY}
The data that support the findings of this study are available from the corresponding author upon reasonable request.

\bibliography{refs}

\begin{thebibliography}{45}%
\makeatletter
\providecommand \@ifxundefined [1]{%
 \@ifx{#1\undefined}
}%
\providecommand \@ifnum [1]{%
 \ifnum #1\expandafter \@firstoftwo
 \else \expandafter \@secondoftwo
 \fi
}%
\providecommand \@ifx [1]{%
 \ifx #1\expandafter \@firstoftwo
 \else \expandafter \@secondoftwo
 \fi
}%
\providecommand \natexlab [1]{#1}%
\providecommand \enquote  [1]{``#1''}%
\providecommand \bibnamefont  [1]{#1}%
\providecommand \bibfnamefont [1]{#1}%
\providecommand \citenamefont [1]{#1}%
\providecommand \href@noop [0]{\@secondoftwo}%
\providecommand \href [0]{\begingroup \@sanitize@url \@href}%
\providecommand \@href[1]{\@@startlink{#1}\@@href}%
\providecommand \@@href[1]{\endgroup#1\@@endlink}%
\providecommand \@sanitize@url [0]{\catcode `\\12\catcode `\$12\catcode
  `\&12\catcode `\#12\catcode `\^12\catcode `\_12\catcode `\%12\relax}%
\providecommand \@@startlink[1]{}%
\providecommand \@@endlink[0]{}%
\providecommand \url  [0]{\begingroup\@sanitize@url \@url }%
\providecommand \@url [1]{\endgroup\@href {#1}{\urlprefix }}%
\providecommand \urlprefix  [0]{URL }%
\providecommand \Eprint [0]{\href }%
\providecommand \doibase [0]{https://doi.org/}%
\providecommand \selectlanguage [0]{\@gobble}%
\providecommand \bibinfo  [0]{\@secondoftwo}%
\providecommand \bibfield  [0]{\@secondoftwo}%
\providecommand \translation [1]{[#1]}%
\providecommand \BibitemOpen [0]{}%
\providecommand \bibitemStop [0]{}%
\providecommand \bibitemNoStop [0]{.\EOS\space}%
\providecommand \EOS [0]{\spacefactor3000\relax}%
\providecommand \BibitemShut  [1]{\csname bibitem#1\endcsname}%
\let\auto@bib@innerbib\@empty
\bibitem [{\citenamefont {Bhatti}\ \emph {et~al.}(2017)\citenamefont {Bhatti},
  \citenamefont {Sbiaa}, \citenamefont {Hirohata}, \citenamefont {Ohno},
  \citenamefont {Fukami},\ and\ \citenamefont
  {Piramanayagam}}]{bhatti2017spintronics}%
  \BibitemOpen
  \bibfield  {author} {\bibinfo {author} {\bibfnamefont {S.}~\bibnamefont
  {Bhatti}}, \bibinfo {author} {\bibfnamefont {R.}~\bibnamefont {Sbiaa}},
  \bibinfo {author} {\bibfnamefont {A.}~\bibnamefont {Hirohata}}, \bibinfo
  {author} {\bibfnamefont {H.}~\bibnamefont {Ohno}}, \bibinfo {author}
  {\bibfnamefont {S.}~\bibnamefont {Fukami}},\ and\ \bibinfo {author}
  {\bibfnamefont {S.}~\bibnamefont {Piramanayagam}},\ }\bibfield  {title}
  {\enquote {\bibinfo {title} {{Spintronics based random access memory: a
  review}},}\ }\href {https://doi.org/10.1016/j.mattod.2017.07.007} {\bibfield
  {journal} {\bibinfo  {journal} {Mater. Today}\ }\textbf {\bibinfo {volume}
  {20}},\ \bibinfo {pages} {530--548} (\bibinfo {year} {2017})}\BibitemShut
  {NoStop}%
\bibitem [{\citenamefont {Hirohata}\ \emph {et~al.}(2020)\citenamefont
  {Hirohata}, \citenamefont {Yamada}, \citenamefont {Nakatani}, \citenamefont
  {Prejbeanu}, \citenamefont {Diény}, \citenamefont {Pirro},\ and\
  \citenamefont {Hillebrands}}]{hirohata2020review}%
  \BibitemOpen
  \bibfield  {author} {\bibinfo {author} {\bibfnamefont {A.}~\bibnamefont
  {Hirohata}}, \bibinfo {author} {\bibfnamefont {K.}~\bibnamefont {Yamada}},
  \bibinfo {author} {\bibfnamefont {Y.}~\bibnamefont {Nakatani}}, \bibinfo
  {author} {\bibfnamefont {I.-L.}\ \bibnamefont {Prejbeanu}}, \bibinfo {author}
  {\bibfnamefont {B.}~\bibnamefont {Diény}}, \bibinfo {author} {\bibfnamefont
  {P.}~\bibnamefont {Pirro}},\ and\ \bibinfo {author} {\bibfnamefont
  {B.}~\bibnamefont {Hillebrands}},\ }\bibfield  {title} {\enquote {\bibinfo
  {title} {Review on spintronics: Principles and device applications},}\ }\href
  {https://doi.org/https://doi.org/10.1016/j.jmmm.2020.166711} {\bibfield
  {journal} {\bibinfo  {journal} {J. Magn. Magn. Mater.}\ }\textbf {\bibinfo
  {volume} {509}},\ \bibinfo {pages} {166711} (\bibinfo {year}
  {2020})}\BibitemShut {NoStop}%
\bibitem [{\citenamefont {Endoh}\ \emph {et~al.}(2020)\citenamefont {Endoh},
  \citenamefont {Honjo}, \citenamefont {Nishioka},\ and\ \citenamefont
  {Ikeda}}]{endoh2020recent}%
  \BibitemOpen
  \bibfield  {author} {\bibinfo {author} {\bibfnamefont {T.}~\bibnamefont
  {Endoh}}, \bibinfo {author} {\bibfnamefont {H.}~\bibnamefont {Honjo}},
  \bibinfo {author} {\bibfnamefont {K.}~\bibnamefont {Nishioka}},\ and\
  \bibinfo {author} {\bibfnamefont {S.}~\bibnamefont {Ikeda}},\ }\bibfield
  {title} {\enquote {\bibinfo {title} {{Recent progresses in STT-MRAM and
  SOT-MRAM for next generation MRAM}},}\ }in\ \href
  {https://doi.org/10.1109/VLSITechnology18217.2020.9265042} {\emph {\bibinfo
  {booktitle} {2020 IEEE Symposium on VLSI Technology}}}\ (\bibinfo
  {organization} {IEEE},\ \bibinfo {year} {2020})\ pp.\ \bibinfo {pages}
  {1--2}\BibitemShut {NoStop}%
\bibitem [{\citenamefont {Naganuma}(2023)}]{naganuma2023spintronics}%
  \BibitemOpen
  \bibfield  {author} {\bibinfo {author} {\bibfnamefont {H.}~\bibnamefont
  {Naganuma}},\ }\bibfield  {title} {\enquote {\bibinfo {title} {Spintronics
  memory using magnetic tunnel junction for {$X$} nm-generation},}\ }\href
  {https://doi.org/10.35848/1347-4065/accaed} {\bibfield  {journal} {\bibinfo
  {journal} {Jpn. J. Appl. Phys.}\ }\textbf {\bibinfo {volume} {62}},\ \bibinfo
  {pages} {SG0811} (\bibinfo {year} {2023})}\BibitemShut {NoStop}%
\bibitem [{\citenamefont {Naganuma}\ \emph {et~al.}(2015)\citenamefont
  {Naganuma}, \citenamefont {Kim}, \citenamefont {Kawada}, \citenamefont
  {Inami}, \citenamefont {Hatakeyama}, \citenamefont {Iihama}, \citenamefont
  {Nazrul~Islam}, \citenamefont {Oogane}, \citenamefont {Mizukami},\ and\
  \citenamefont {Ando}}]{naganuma2015electrical}%
  \BibitemOpen
  \bibfield  {author} {\bibinfo {author} {\bibfnamefont {H.}~\bibnamefont
  {Naganuma}}, \bibinfo {author} {\bibfnamefont {G.}~\bibnamefont {Kim}},
  \bibinfo {author} {\bibfnamefont {Y.}~\bibnamefont {Kawada}}, \bibinfo
  {author} {\bibfnamefont {N.}~\bibnamefont {Inami}}, \bibinfo {author}
  {\bibfnamefont {K.}~\bibnamefont {Hatakeyama}}, \bibinfo {author}
  {\bibfnamefont {S.}~\bibnamefont {Iihama}}, \bibinfo {author} {\bibfnamefont
  {K.~M.}\ \bibnamefont {Nazrul~Islam}}, \bibinfo {author} {\bibfnamefont
  {M.}~\bibnamefont {Oogane}}, \bibinfo {author} {\bibfnamefont
  {S.}~\bibnamefont {Mizukami}},\ and\ \bibinfo {author} {\bibfnamefont
  {Y.}~\bibnamefont {Ando}},\ }\bibfield  {title} {\enquote {\bibinfo {title}
  {{Electrical detection of millimeter-waves by magnetic tunnel junctions using
  perpendicular magnetized $L1_0$-FePd free layer}},}\ }\href
  {https://doi.org/10.1021/nl504114v} {\bibfield  {journal} {\bibinfo
  {journal} {Nano Lett.}\ }\textbf {\bibinfo {volume} {15}},\ \bibinfo {pages}
  {623--628} (\bibinfo {year} {2015})}\BibitemShut {NoStop}%
\bibitem [{\citenamefont {Naganuma}\ \emph {et~al.}(2020)\citenamefont
  {Naganuma}, \citenamefont {Zatko}, \citenamefont {Galbiati}, \citenamefont
  {Godel}, \citenamefont {Sander}, \citenamefont {Carr{\'e}t{\'e}ro},
  \citenamefont {Bezencenet}, \citenamefont {Reyren}, \citenamefont {Martin},
  \citenamefont {Dlubak},\ and\ \citenamefont
  {Seneor}}]{naganuma2020perpendicular}%
  \BibitemOpen
  \bibfield  {author} {\bibinfo {author} {\bibfnamefont {H.}~\bibnamefont
  {Naganuma}}, \bibinfo {author} {\bibfnamefont {V.}~\bibnamefont {Zatko}},
  \bibinfo {author} {\bibfnamefont {M.}~\bibnamefont {Galbiati}}, \bibinfo
  {author} {\bibfnamefont {F.}~\bibnamefont {Godel}}, \bibinfo {author}
  {\bibfnamefont {A.}~\bibnamefont {Sander}}, \bibinfo {author} {\bibfnamefont
  {C.}~\bibnamefont {Carr{\'e}t{\'e}ro}}, \bibinfo {author} {\bibfnamefont
  {O.}~\bibnamefont {Bezencenet}}, \bibinfo {author} {\bibfnamefont
  {N.}~\bibnamefont {Reyren}}, \bibinfo {author} {\bibfnamefont {M.-B.}\
  \bibnamefont {Martin}}, \bibinfo {author} {\bibfnamefont {B.}~\bibnamefont
  {Dlubak}},\ and\ \bibinfo {author} {\bibfnamefont {P.}~\bibnamefont
  {Seneor}},\ }\bibfield  {title} {\enquote {\bibinfo {title} {{A perpendicular
  graphene/ferromagnet electrode for spintronics}},}\ }\href
  {https://doi.org/10.1063/1.5143567} {\bibfield  {journal} {\bibinfo
  {journal} {Appl. Phys. Lett.}\ }\textbf {\bibinfo {volume} {116}},\ \bibinfo
  {pages} {173101} (\bibinfo {year} {2020})}\BibitemShut {NoStop}%
\bibitem [{\citenamefont {Naganuma}\ \emph {et~al.}(2022)\citenamefont
  {Naganuma}, \citenamefont {Nishijima}, \citenamefont {Adachi}, \citenamefont
  {Uemoto}, \citenamefont {Shinya}, \citenamefont {Yasui}, \citenamefont
  {Morioka}, \citenamefont {Hirata}, \citenamefont {Godel}, \citenamefont
  {Martin}, \citenamefont {Dlubak}, \citenamefont {Seneor},\ and\ \citenamefont
  {Amemiya}}]{naganuma2022unveiling}%
  \BibitemOpen
  \bibfield  {author} {\bibinfo {author} {\bibfnamefont {H.}~\bibnamefont
  {Naganuma}}, \bibinfo {author} {\bibfnamefont {M.}~\bibnamefont {Nishijima}},
  \bibinfo {author} {\bibfnamefont {H.}~\bibnamefont {Adachi}}, \bibinfo
  {author} {\bibfnamefont {M.}~\bibnamefont {Uemoto}}, \bibinfo {author}
  {\bibfnamefont {H.}~\bibnamefont {Shinya}}, \bibinfo {author} {\bibfnamefont
  {S.}~\bibnamefont {Yasui}}, \bibinfo {author} {\bibfnamefont
  {H.}~\bibnamefont {Morioka}}, \bibinfo {author} {\bibfnamefont
  {A.}~\bibnamefont {Hirata}}, \bibinfo {author} {\bibfnamefont
  {F.}~\bibnamefont {Godel}}, \bibinfo {author} {\bibfnamefont {M.-B.}\
  \bibnamefont {Martin}}, \bibinfo {author} {\bibfnamefont {B.}~\bibnamefont
  {Dlubak}}, \bibinfo {author} {\bibfnamefont {P.}~\bibnamefont {Seneor}},\
  and\ \bibinfo {author} {\bibfnamefont {K.}~\bibnamefont {Amemiya}},\
  }\bibfield  {title} {\enquote {\bibinfo {title} {Unveiling a chemisorbed
  crystallographically heterogeneous graphene/{$L1_0$-FePd} interface with a
  robust and perpendicular orbital moment},}\ }\href
  {https://doi.org/10.1021/acsnano.1c09843} {\bibfield  {journal} {\bibinfo
  {journal} {ACS nano}\ }\textbf {\bibinfo {volume} {16}},\ \bibinfo {pages}
  {4139--4151} (\bibinfo {year} {2022})}\BibitemShut {NoStop}%
\bibitem [{\citenamefont {Uemoto}\ \emph {et~al.}(2022)\citenamefont {Uemoto},
  \citenamefont {Adachi}, \citenamefont {Naganuma},\ and\ \citenamefont
  {Ono}}]{uemoto2022density}%
  \BibitemOpen
  \bibfield  {author} {\bibinfo {author} {\bibfnamefont {M.}~\bibnamefont
  {Uemoto}}, \bibinfo {author} {\bibfnamefont {H.}~\bibnamefont {Adachi}},
  \bibinfo {author} {\bibfnamefont {H.}~\bibnamefont {Naganuma}},\ and\
  \bibinfo {author} {\bibfnamefont {T.}~\bibnamefont {Ono}},\ }\bibfield
  {title} {\enquote {\bibinfo {title} {Density functional study of twisted
  graphene {$L1_0$-FePd} heterogeneous interface},}\ }\href
  {https://doi.org/10.1063/5.0101703} {\bibfield  {journal} {\bibinfo
  {journal} {J. Appl. Phys.}\ }\textbf {\bibinfo {volume} {132}},\ \bibinfo
  {pages} {095301} (\bibinfo {year} {2022})}\BibitemShut {NoStop}%
\bibitem [{\citenamefont {Naganuma}\ \emph {et~al.}(2023)\citenamefont
  {Naganuma}, \citenamefont {Uemoto}, \citenamefont {Adachi}, \citenamefont
  {Shinya}, \citenamefont {Mochizuki}, \citenamefont {Kobayashi}, \citenamefont
  {Hirata}, \citenamefont {Dlubak}, \citenamefont {Ono}, \citenamefont
  {Seneor}, \citenamefont {Robertson},\ and\ \citenamefont
  {Amemiya}}]{naganuma2023jpcc}%
  \BibitemOpen
  \bibfield  {author} {\bibinfo {author} {\bibfnamefont {H.}~\bibnamefont
  {Naganuma}}, \bibinfo {author} {\bibfnamefont {M.}~\bibnamefont {Uemoto}},
  \bibinfo {author} {\bibfnamefont {H.}~\bibnamefont {Adachi}}, \bibinfo
  {author} {\bibfnamefont {H.}~\bibnamefont {Shinya}}, \bibinfo {author}
  {\bibfnamefont {I.}~\bibnamefont {Mochizuki}}, \bibinfo {author}
  {\bibfnamefont {M.}~\bibnamefont {Kobayashi}}, \bibinfo {author}
  {\bibfnamefont {A.}~\bibnamefont {Hirata}}, \bibinfo {author} {\bibfnamefont
  {B.}~\bibnamefont {Dlubak}}, \bibinfo {author} {\bibfnamefont
  {T.}~\bibnamefont {Ono}}, \bibinfo {author} {\bibfnamefont {P.}~\bibnamefont
  {Seneor}}, \bibinfo {author} {\bibfnamefont {J.}~\bibnamefont {Robertson}},\
  and\ \bibinfo {author} {\bibfnamefont {K.}~\bibnamefont {Amemiya}},\
  }\bibfield  {title} {\enquote {\bibinfo {title} {Twist pz orbital and spin
  moment of the wavy-graphene/{$L1_0$}-fepd moiré interface},}\ }\href
  {https://doi.org/10.1021/acs.jpcc.2c08982} {\bibfield  {journal} {\bibinfo
  {journal} {J. Phys. Chem. C}\ }\textbf {\bibinfo {volume} {127}},\ \bibinfo
  {pages} {11481--11489} (\bibinfo {year} {2023})}\BibitemShut {NoStop}%
\bibitem [{\citenamefont {Zhang}\ \emph {et~al.}(2018)\citenamefont {Zhang},
  \citenamefont {Schliep}, \citenamefont {Wu}, \citenamefont {Quarterman},
  \citenamefont {Reifsnyder~Hickey}, \citenamefont {Lv}, \citenamefont {Chao},
  \citenamefont {Li}, \citenamefont {Chen}, \citenamefont {Zhao}, \citenamefont
  {Jamali}, \citenamefont {Mkhoya},\ and\ \citenamefont
  {Wang}}]{zhang2018enhancement}%
  \BibitemOpen
  \bibfield  {author} {\bibinfo {author} {\bibfnamefont {D.-L.}\ \bibnamefont
  {Zhang}}, \bibinfo {author} {\bibfnamefont {K.~B.}\ \bibnamefont {Schliep}},
  \bibinfo {author} {\bibfnamefont {R.~J.}\ \bibnamefont {Wu}}, \bibinfo
  {author} {\bibfnamefont {P.}~\bibnamefont {Quarterman}}, \bibinfo {author}
  {\bibfnamefont {D.}~\bibnamefont {Reifsnyder~Hickey}}, \bibinfo {author}
  {\bibfnamefont {Y.}~\bibnamefont {Lv}}, \bibinfo {author} {\bibfnamefont
  {X.}~\bibnamefont {Chao}}, \bibinfo {author} {\bibfnamefont {H.}~\bibnamefont
  {Li}}, \bibinfo {author} {\bibfnamefont {J.-Y.}\ \bibnamefont {Chen}},
  \bibinfo {author} {\bibfnamefont {Z.}~\bibnamefont {Zhao}}, \bibinfo {author}
  {\bibfnamefont {M.}~\bibnamefont {Jamali}}, \bibinfo {author} {\bibfnamefont
  {A.~K.}\ \bibnamefont {Mkhoya}},\ and\ \bibinfo {author} {\bibfnamefont
  {J.-P.}\ \bibnamefont {Wang}},\ }\bibfield  {title} {\enquote {\bibinfo
  {title} {Enhancement of tunneling magnetoresistance by inserting a diffusion
  barrier in {$L1_0$-FePd} perpendicular magnetic tunnel junctions},}\ }\href
  {https://doi.org/10.1063/1.5019193} {\bibfield  {journal} {\bibinfo
  {journal} {Appl. Phys. Lett.}\ }\textbf {\bibinfo {volume} {112}},\ \bibinfo
  {pages} {152401} (\bibinfo {year} {2018})}\BibitemShut {NoStop}%
\bibitem [{\citenamefont {Mohri}\ \emph {et~al.}(2001)\citenamefont {Mohri},
  \citenamefont {Horiuchi}, \citenamefont {Uzawa}, \citenamefont {Ibaragi},
  \citenamefont {Igarashi},\ and\ \citenamefont {Abe}}]{mohri2001theoretical}%
  \BibitemOpen
  \bibfield  {author} {\bibinfo {author} {\bibfnamefont {T.}~\bibnamefont
  {Mohri}}, \bibinfo {author} {\bibfnamefont {T.}~\bibnamefont {Horiuchi}},
  \bibinfo {author} {\bibfnamefont {H.}~\bibnamefont {Uzawa}}, \bibinfo
  {author} {\bibfnamefont {M.}~\bibnamefont {Ibaragi}}, \bibinfo {author}
  {\bibfnamefont {M.}~\bibnamefont {Igarashi}},\ and\ \bibinfo {author}
  {\bibfnamefont {F.}~\bibnamefont {Abe}},\ }\bibfield  {title} {\enquote
  {\bibinfo {title} {{Theoretical investigation of $L1_0$-disorder phase
  equilibria in Fe--Pd alloy system}},}\ }\href
  {https://doi.org/10.1016/S0925-8388(00)01408-0} {\bibfield  {journal}
  {\bibinfo  {journal} {J. Alloys Compd.}\ }\textbf {\bibinfo {volume} {317}},\
  \bibinfo {pages} {13--18} (\bibinfo {year} {2001})}\BibitemShut {NoStop}%
\bibitem [{\citenamefont {Klemmer}\ \emph {et~al.}(1995)\citenamefont
  {Klemmer}, \citenamefont {Hoydick}, \citenamefont {Okumura}, \citenamefont
  {Zhang},\ and\ \citenamefont {Soffa}}]{klemmer1995magnetic}%
  \BibitemOpen
  \bibfield  {author} {\bibinfo {author} {\bibfnamefont {T.}~\bibnamefont
  {Klemmer}}, \bibinfo {author} {\bibfnamefont {D.}~\bibnamefont {Hoydick}},
  \bibinfo {author} {\bibfnamefont {H.}~\bibnamefont {Okumura}}, \bibinfo
  {author} {\bibfnamefont {B.}~\bibnamefont {Zhang}},\ and\ \bibinfo {author}
  {\bibfnamefont {W.}~\bibnamefont {Soffa}},\ }\bibfield  {title} {\enquote
  {\bibinfo {title} {{Magnetic hardening and coercivity mechanisms in $L1_0$
  ordered FePd ferromagnets}},}\ }\href
  {https://doi.org/10.1016/0956-716X(95)00413-P} {\bibfield  {journal}
  {\bibinfo  {journal} {Scr. Mater.}\ }\textbf {\bibinfo {volume} {33}},\
  \bibinfo {pages} {1793--1805} (\bibinfo {year} {1995})}\BibitemShut {NoStop}%
\bibitem [{\citenamefont {Shima}\ \emph {et~al.}(2004)\citenamefont {Shima},
  \citenamefont {Oikawa}, \citenamefont {Fujita}, \citenamefont {Fukamichi},
  \citenamefont {Ishida},\ and\ \citenamefont {Sakuma}}]{shima2004lattice}%
  \BibitemOpen
  \bibfield  {author} {\bibinfo {author} {\bibfnamefont {H.}~\bibnamefont
  {Shima}}, \bibinfo {author} {\bibfnamefont {K.}~\bibnamefont {Oikawa}},
  \bibinfo {author} {\bibfnamefont {A.}~\bibnamefont {Fujita}}, \bibinfo
  {author} {\bibfnamefont {K.}~\bibnamefont {Fukamichi}}, \bibinfo {author}
  {\bibfnamefont {K.}~\bibnamefont {Ishida}},\ and\ \bibinfo {author}
  {\bibfnamefont {A.}~\bibnamefont {Sakuma}},\ }\bibfield  {title} {\enquote
  {\bibinfo {title} {{Lattice axial ratio and large uniaxial magnetocrystalline
  anisotropy in $L{1}_{0}$-type FePd single crystals prepared under compressive
  stress}},}\ }\href {https://doi.org/10.1103/PhysRevB.70.224408} {\bibfield
  {journal} {\bibinfo  {journal} {Phys. Rev. B}\ }\textbf {\bibinfo {volume}
  {70}},\ \bibinfo {pages} {224408} (\bibinfo {year} {2004})}\BibitemShut
  {NoStop}%
\bibitem [{\citenamefont {Miyata}\ \emph {et~al.}(1990)\citenamefont {Miyata},
  \citenamefont {Asami}, \citenamefont {Mizushima},\ and\ \citenamefont
  {Sato}}]{miyata1990ferromagnetic}%
  \BibitemOpen
  \bibfield  {author} {\bibinfo {author} {\bibfnamefont {N.}~\bibnamefont
  {Miyata}}, \bibinfo {author} {\bibfnamefont {H.}~\bibnamefont {Asami}},
  \bibinfo {author} {\bibfnamefont {T.}~\bibnamefont {Mizushima}},\ and\
  \bibinfo {author} {\bibfnamefont {K.}~\bibnamefont {Sato}},\ }\bibfield
  {title} {\enquote {\bibinfo {title} {{Ferromagnetic Crystalline Anisotropy of
  Pd${}_{1-x}$ Fe${}_x$ Alloys. III. $0.38 \sim 0.5$, $L1_0$-Type Ordered
  Phase}},}\ }\href {https://doi.org/10.1143/JPSJ.59.1817} {\bibfield
  {journal} {\bibinfo  {journal} {J. Phys. Soc. Jpn.}\ }\textbf {\bibinfo
  {volume} {59}},\ \bibinfo {pages} {1817--1824} (\bibinfo {year}
  {1990})}\BibitemShut {NoStop}%
\bibitem [{\citenamefont {Iihama}\ \emph {et~al.}(2014)\citenamefont {Iihama},
  \citenamefont {Sakuma}, \citenamefont {Naganuma}, \citenamefont {Oogane},
  \citenamefont {Miyazaki}, \citenamefont {Mizukami},\ and\ \citenamefont
  {Ando}}]{iihama2014low}%
  \BibitemOpen
  \bibfield  {author} {\bibinfo {author} {\bibfnamefont {S.}~\bibnamefont
  {Iihama}}, \bibinfo {author} {\bibfnamefont {A.}~\bibnamefont {Sakuma}},
  \bibinfo {author} {\bibfnamefont {H.}~\bibnamefont {Naganuma}}, \bibinfo
  {author} {\bibfnamefont {M.}~\bibnamefont {Oogane}}, \bibinfo {author}
  {\bibfnamefont {T.}~\bibnamefont {Miyazaki}}, \bibinfo {author}
  {\bibfnamefont {S.}~\bibnamefont {Mizukami}},\ and\ \bibinfo {author}
  {\bibfnamefont {Y.}~\bibnamefont {Ando}},\ }\bibfield  {title} {\enquote
  {\bibinfo {title} {{Low precessional damping observed for $L1_0$-ordered FePd
  epitaxial thin films with large perpendicular magnetic anisotropy}},}\ }\href
  {https://doi.org/10.1063/1.4897547} {\bibfield  {journal} {\bibinfo
  {journal} {Appl. Phys. Lett.}\ }\textbf {\bibinfo {volume} {105}},\ \bibinfo
  {pages} {142403} (\bibinfo {year} {2014})}\BibitemShut {NoStop}%
\bibitem [{\citenamefont {Kawai}\ \emph {et~al.}(2014)\citenamefont {Kawai},
  \citenamefont {Itabashi}, \citenamefont {Ohtake}, \citenamefont {Takeda},\
  and\ \citenamefont {Futamoto}}]{kawai2014gilbert}%
  \BibitemOpen
  \bibfield  {author} {\bibinfo {author} {\bibfnamefont {T.}~\bibnamefont
  {Kawai}}, \bibinfo {author} {\bibfnamefont {A.}~\bibnamefont {Itabashi}},
  \bibinfo {author} {\bibfnamefont {M.}~\bibnamefont {Ohtake}}, \bibinfo
  {author} {\bibfnamefont {S.}~\bibnamefont {Takeda}},\ and\ \bibinfo {author}
  {\bibfnamefont {M.}~\bibnamefont {Futamoto}},\ }\bibfield  {title} {\enquote
  {\bibinfo {title} {{Gilbert damping constant of FePd alloy thin films
  estimated by broadband ferromagnetic resonance}},}\ }in\ \href
  {https://doi.org/10.1051/epjconf/20147502002} {\emph {\bibinfo {booktitle}
  {{EPJ Web of Conferences}}}},\ Vol.~\bibinfo {volume} {75}\ (\bibinfo
  {organization} {EDP Sciences},\ \bibinfo {year} {2014})\ p.\ \bibinfo {pages}
  {02002}\BibitemShut {NoStop}%
\bibitem [{\citenamefont {Itabashi}\ \emph {et~al.}(2013)\citenamefont
  {Itabashi}, \citenamefont {Ohtake}, \citenamefont {Ouchi}, \citenamefont
  {Kirino},\ and\ \citenamefont {Futamoto}}]{itabashi2013preparation}%
  \BibitemOpen
  \bibfield  {author} {\bibinfo {author} {\bibfnamefont {A.}~\bibnamefont
  {Itabashi}}, \bibinfo {author} {\bibfnamefont {M.}~\bibnamefont {Ohtake}},
  \bibinfo {author} {\bibfnamefont {S.}~\bibnamefont {Ouchi}}, \bibinfo
  {author} {\bibfnamefont {F.}~\bibnamefont {Kirino}},\ and\ \bibinfo {author}
  {\bibfnamefont {M.}~\bibnamefont {Futamoto}},\ }\bibfield  {title} {\enquote
  {\bibinfo {title} {{Preparation of $L1_0$ ordered FePd, FePt, and CoPt thin
  films with flat surfaces on MgO (001) single-crystal substrates}},}\ }in\
  \href {https://doi.org/10.1051/epjconf/20134007001} {\emph {\bibinfo
  {booktitle} {{EPJ Web of Conferences}}}},\ Vol.~\bibinfo {volume} {40}\
  (\bibinfo {organization} {EDP Sciences},\ \bibinfo {year} {2013})\ p.\
  \bibinfo {pages} {07001}\BibitemShut {NoStop}%
\bibitem [{\citenamefont {Hallal}\ \emph {et~al.}(2013)\citenamefont {Hallal},
  \citenamefont {Yang}, \citenamefont {Dieny},\ and\ \citenamefont
  {Chshiev}}]{hallal2013anatomy}%
  \BibitemOpen
  \bibfield  {author} {\bibinfo {author} {\bibfnamefont {A.}~\bibnamefont
  {Hallal}}, \bibinfo {author} {\bibfnamefont {H.}~\bibnamefont {Yang}},
  \bibinfo {author} {\bibfnamefont {B.}~\bibnamefont {Dieny}},\ and\ \bibinfo
  {author} {\bibfnamefont {M.}~\bibnamefont {Chshiev}},\ }\bibfield  {title}
  {\enquote {\bibinfo {title} {Anatomy of perpendicular magnetic anisotropy in
  fe/mgo magnetic tunnel junctions: First-principles insight},}\ }\href
  {https://doi.org/10.1103/PhysRevB.88.184423} {\bibfield  {journal} {\bibinfo
  {journal} {Phys. Rev. B}\ }\textbf {\bibinfo {volume} {88}},\ \bibinfo
  {pages} {184423} (\bibinfo {year} {2013})}\BibitemShut {NoStop}%
\bibitem [{\citenamefont {Lu}, \citenamefont {Robertson},\ and\ \citenamefont
  {Naganuma}(2021)}]{lu2021comparison}%
  \BibitemOpen
  \bibfield  {author} {\bibinfo {author} {\bibfnamefont {H.}~\bibnamefont
  {Lu}}, \bibinfo {author} {\bibfnamefont {J.}~\bibnamefont {Robertson}},\ and\
  \bibinfo {author} {\bibfnamefont {H.}~\bibnamefont {Naganuma}},\ }\bibfield
  {title} {\enquote {\bibinfo {title} {{Comparison of hexagonal boron nitride
  and MgO tunnel barriers in Fe,Co magnetic tunnel junctions}},}\ }\href
  {https://doi.org/10.1063/5.0049792} {\bibfield  {journal} {\bibinfo
  {journal} {Appl. Phys. Rev.}\ }\textbf {\bibinfo {volume} {8}},\ \bibinfo
  {pages} {031307} (\bibinfo {year} {2021})}\BibitemShut {NoStop}%
\bibitem [{\citenamefont {Robertson}, \citenamefont {Naganuma},\ and\
  \citenamefont {Lu}(2023)}]{robertson2023comparing}%
  \BibitemOpen
  \bibfield  {author} {\bibinfo {author} {\bibfnamefont {J.}~\bibnamefont
  {Robertson}}, \bibinfo {author} {\bibfnamefont {H.}~\bibnamefont
  {Naganuma}},\ and\ \bibinfo {author} {\bibfnamefont {H.}~\bibnamefont {Lu}},\
  }\bibfield  {title} {\enquote {\bibinfo {title} {Comparing {h-BN} and {MgO}
  tunnel barriers for scaled magnetic tunnel junctions},}\ }\href
  {https://doi.org/10.35848/1347-4065/acb062} {\bibfield  {journal} {\bibinfo
  {journal} {Jpn. J. Appl. Phys.}\ }\textbf {\bibinfo {volume} {62}},\ \bibinfo
  {pages} {SC0804} (\bibinfo {year} {2023})}\BibitemShut {NoStop}%
\bibitem [{\citenamefont {Parkin}\ \emph {et~al.}(2004)\citenamefont {Parkin},
  \citenamefont {Kaiser}, \citenamefont {Panchula}, \citenamefont {Rice},
  \citenamefont {Hughes}, \citenamefont {Samant},\ and\ \citenamefont
  {Yang}}]{parkin2004giant}%
  \BibitemOpen
  \bibfield  {author} {\bibinfo {author} {\bibfnamefont {S.~S.}\ \bibnamefont
  {Parkin}}, \bibinfo {author} {\bibfnamefont {C.}~\bibnamefont {Kaiser}},
  \bibinfo {author} {\bibfnamefont {A.}~\bibnamefont {Panchula}}, \bibinfo
  {author} {\bibfnamefont {P.~M.}\ \bibnamefont {Rice}}, \bibinfo {author}
  {\bibfnamefont {B.}~\bibnamefont {Hughes}}, \bibinfo {author} {\bibfnamefont
  {M.}~\bibnamefont {Samant}},\ and\ \bibinfo {author} {\bibfnamefont {S.-H.}\
  \bibnamefont {Yang}},\ }\bibfield  {title} {\enquote {\bibinfo {title}
  {{Giant tunnelling magnetoresistance at room temperature with MgO (100)
  tunnel barriers}},}\ }\href {https://doi.org/10.1038/nmat1257} {\bibfield
  {journal} {\bibinfo  {journal} {{Nat. Mat.}}\ }\textbf {\bibinfo {volume}
  {3}},\ \bibinfo {pages} {862--867} (\bibinfo {year} {2004})}\BibitemShut
  {NoStop}%
\bibitem [{\citenamefont {Lu}, \citenamefont {Guo},\ and\ \citenamefont
  {Robertson}(2021)}]{lu2021ab}%
  \BibitemOpen
  \bibfield  {author} {\bibinfo {author} {\bibfnamefont {H.}~\bibnamefont
  {Lu}}, \bibinfo {author} {\bibfnamefont {Y.}~\bibnamefont {Guo}},\ and\
  \bibinfo {author} {\bibfnamefont {J.}~\bibnamefont {Robertson}},\ }\bibfield
  {title} {\enquote {\bibinfo {title} {Ab initio study of hexagonal boron
  nitride as the tunnel barrier in magnetic tunnel junctions},}\ }\href
  {https://doi.org/10.1021/acsami.1c13583} {\bibfield  {journal} {\bibinfo
  {journal} {ACS Applied Materials \& Interfaces}\ }\textbf {\bibinfo {volume}
  {13}},\ \bibinfo {pages} {47226--47235} (\bibinfo {year} {2021})}\BibitemShut
  {NoStop}%
\bibitem [{\citenamefont {Yazyev}\ and\ \citenamefont
  {Pasquarello}(2009)}]{yazyev2009magnetoresistive}%
  \BibitemOpen
  \bibfield  {author} {\bibinfo {author} {\bibfnamefont {O.~V.}\ \bibnamefont
  {Yazyev}}\ and\ \bibinfo {author} {\bibfnamefont {A.}~\bibnamefont
  {Pasquarello}},\ }\bibfield  {title} {\enquote {\bibinfo {title}
  {Magnetoresistive junctions based on epitaxial graphene and hexagonal boron
  nitride},}\ }\href {https://doi.org/10.1103/PhysRevB.80.035408} {\bibfield
  {journal} {\bibinfo  {journal} {Phys. Rev. B}\ }\textbf {\bibinfo {volume}
  {80}},\ \bibinfo {pages} {035408} (\bibinfo {year} {2009})}\BibitemShut
  {NoStop}%
\bibitem [{\citenamefont {Luan}\ \emph {et~al.}(2015)\citenamefont {Luan},
  \citenamefont {Zhang}, \citenamefont {Jiao},\ and\ \citenamefont
  {Sun}}]{luan2015tunneling}%
  \BibitemOpen
  \bibfield  {author} {\bibinfo {author} {\bibfnamefont {G.-P.}\ \bibnamefont
  {Luan}}, \bibinfo {author} {\bibfnamefont {P.-R.}\ \bibnamefont {Zhang}},
  \bibinfo {author} {\bibfnamefont {N.}~\bibnamefont {Jiao}},\ and\ \bibinfo
  {author} {\bibfnamefont {L.-Z.}\ \bibnamefont {Sun}},\ }\bibfield  {title}
  {\enquote {\bibinfo {title} {Tunneling magnetoresistance based on a
  cr/graphene/cr magnetotunnel junction},}\ }\href
  {https://doi.org/http://dx.doi.org/10.1088/1674-1056/24/11/117201} {\bibfield
   {journal} {\bibinfo  {journal} {Chin. Phys. B}\ }\textbf {\bibinfo {volume}
  {24}},\ \bibinfo {pages} {117201} (\bibinfo {year} {2015})}\BibitemShut
  {NoStop}%
\bibitem [{\citenamefont {Iqbal}, \citenamefont {Qureshi},\ and\ \citenamefont
  {Hussain}(2018)}]{iqbal2018recent}%
  \BibitemOpen
  \bibfield  {author} {\bibinfo {author} {\bibfnamefont {M.~Z.}\ \bibnamefont
  {Iqbal}}, \bibinfo {author} {\bibfnamefont {N.~A.}\ \bibnamefont {Qureshi}},\
  and\ \bibinfo {author} {\bibfnamefont {G.}~\bibnamefont {Hussain}},\
  }\bibfield  {title} {\enquote {\bibinfo {title} {Recent advancements in
  {2D}-materials interface based magnetic junctions for spintronics},}\ }\href
  {https://doi.org/10.1016/j.jmmm.2018.02.084} {\bibfield  {journal} {\bibinfo
  {journal} {J. Magn. Magn. Mater.}\ }\textbf {\bibinfo {volume} {457}},\
  \bibinfo {pages} {110--125} (\bibinfo {year} {2018})}\BibitemShut {NoStop}%
\bibitem [{\citenamefont {Hashmi}, \citenamefont {Nakanishi},\ and\
  \citenamefont {Ono}(2020)}]{hashmi2020graphene}%
  \BibitemOpen
  \bibfield  {author} {\bibinfo {author} {\bibfnamefont {A.}~\bibnamefont
  {Hashmi}}, \bibinfo {author} {\bibfnamefont {K.}~\bibnamefont {Nakanishi}},\
  and\ \bibinfo {author} {\bibfnamefont {T.}~\bibnamefont {Ono}},\ }\bibfield
  {title} {\enquote {\bibinfo {title} {Graphene-based symmetric and
  non-symmetric magnetoresistive junctions},}\ }\href
  {https://doi.org/10.7566/JPSJ.89.034708} {\bibfield  {journal} {\bibinfo
  {journal} {J. Phys. Soc. Jpn.}\ }\textbf {\bibinfo {volume} {89}},\ \bibinfo
  {pages} {034708} (\bibinfo {year} {2020})}\BibitemShut {NoStop}%
\bibitem [{\citenamefont {Piquemal-Banci}\ \emph {et~al.}(2020)\citenamefont
  {Piquemal-Banci}, \citenamefont {Galceran}, \citenamefont {Dubois},
  \citenamefont {Zatko}, \citenamefont {Galbiati}, \citenamefont {Godel},
  \citenamefont {Martin}, \citenamefont {Weatherup}, \citenamefont {Petroff},
  \citenamefont {Fert}, \citenamefont {Charlier}, \citenamefont {Robertson},
  \citenamefont {Hofmann}, \citenamefont {Dlubak},\ and\ \citenamefont
  {Seneor}}]{piquemal2020spin}%
  \BibitemOpen
  \bibfield  {author} {\bibinfo {author} {\bibfnamefont {M.}~\bibnamefont
  {Piquemal-Banci}}, \bibinfo {author} {\bibfnamefont {R.}~\bibnamefont
  {Galceran}}, \bibinfo {author} {\bibfnamefont {S.~M.-M.}\ \bibnamefont
  {Dubois}}, \bibinfo {author} {\bibfnamefont {V.}~\bibnamefont {Zatko}},
  \bibinfo {author} {\bibfnamefont {M.}~\bibnamefont {Galbiati}}, \bibinfo
  {author} {\bibfnamefont {F.}~\bibnamefont {Godel}}, \bibinfo {author}
  {\bibfnamefont {M.-B.}\ \bibnamefont {Martin}}, \bibinfo {author}
  {\bibfnamefont {R.~S.}\ \bibnamefont {Weatherup}}, \bibinfo {author}
  {\bibfnamefont {F.}~\bibnamefont {Petroff}}, \bibinfo {author} {\bibfnamefont
  {A.}~\bibnamefont {Fert}}, \bibinfo {author} {\bibfnamefont {J.-C.}\
  \bibnamefont {Charlier}}, \bibinfo {author} {\bibfnamefont {J.}~\bibnamefont
  {Robertson}}, \bibinfo {author} {\bibfnamefont {S.}~\bibnamefont {Hofmann}},
  \bibinfo {author} {\bibfnamefont {B.}~\bibnamefont {Dlubak}},\ and\ \bibinfo
  {author} {\bibfnamefont {P.}~\bibnamefont {Seneor}},\ }\bibfield  {title}
  {\enquote {\bibinfo {title} {{Spin filtering by proximity effects at
  hybridized interfaces in spin-valves with 2D graphene barriers}},}\ }\href
  {https://doi.org/10.1038/s41467-020-19420-6} {\bibfield  {journal} {\bibinfo
  {journal} {Nat. Commun.}\ }\textbf {\bibinfo {volume} {11}},\ \bibinfo
  {pages} {1--9} (\bibinfo {year} {2020})}\BibitemShut {NoStop}%
\bibitem [{\citenamefont {Ahn}(2020)}]{ahn20202d}%
  \BibitemOpen
  \bibfield  {author} {\bibinfo {author} {\bibfnamefont {E.~C.}\ \bibnamefont
  {Ahn}},\ }\bibfield  {title} {\enquote {\bibinfo {title} {{2D} materials for
  spintronic devices},}\ }\href {https://doi.org/10.1038/s41699-020-0152-0}
  {\bibfield  {journal} {\bibinfo  {journal} {npj 2D Mater Appl}\ }\textbf
  {\bibinfo {volume} {4}},\ \bibinfo {pages} {17} (\bibinfo {year}
  {2020})}\BibitemShut {NoStop}%
\bibitem [{\citenamefont {Avsar}\ \emph {et~al.}(2020)\citenamefont {Avsar},
  \citenamefont {Ochoa}, \citenamefont {Guinea}, \citenamefont {\"Ozyilmaz},
  \citenamefont {van Wees},\ and\ \citenamefont
  {Vera-Marun}}]{Avsar2020review}%
  \BibitemOpen
  \bibfield  {author} {\bibinfo {author} {\bibfnamefont {A.}~\bibnamefont
  {Avsar}}, \bibinfo {author} {\bibfnamefont {H.}~\bibnamefont {Ochoa}},
  \bibinfo {author} {\bibfnamefont {F.}~\bibnamefont {Guinea}}, \bibinfo
  {author} {\bibfnamefont {B.}~\bibnamefont {\"Ozyilmaz}}, \bibinfo {author}
  {\bibfnamefont {B.~J.}\ \bibnamefont {van Wees}},\ and\ \bibinfo {author}
  {\bibfnamefont {I.~J.}\ \bibnamefont {Vera-Marun}},\ }\bibfield  {title}
  {\enquote {\bibinfo {title} {Colloquium: Spintronics in graphene and other
  two-dimensional materials},}\ }\href
  {https://doi.org/10.1103/RevModPhys.92.021003} {\bibfield  {journal}
  {\bibinfo  {journal} {Rev. Mod. Phys.}\ }\textbf {\bibinfo {volume} {92}},\
  \bibinfo {pages} {021003} (\bibinfo {year} {2020})}\BibitemShut {NoStop}%
\bibitem [{\citenamefont {Kozlov}, \citenamefont {Vines},\ and\ \citenamefont
  {G\"orling}(2012)}]{kozlov2012bonding}%
  \BibitemOpen
  \bibfield  {author} {\bibinfo {author} {\bibfnamefont {S.~M.}\ \bibnamefont
  {Kozlov}}, \bibinfo {author} {\bibfnamefont {F.}~\bibnamefont {Vines}},\ and\
  \bibinfo {author} {\bibfnamefont {A.}~\bibnamefont {G\"orling}},\ }\bibfield
  {title} {\enquote {\bibinfo {title} {{Bonding Mechanisms of Graphene on Metal
  Surfaces}},}\ }\href {https://doi.org/10.1021/jp210667f} {\bibfield
  {journal} {\bibinfo  {journal} {J. Phys. Chem. C}\ }\textbf {\bibinfo
  {volume} {116}},\ \bibinfo {pages} {7360--7366} (\bibinfo {year}
  {2012})}\BibitemShut {NoStop}%
\bibitem [{\citenamefont {Ono}\ and\ \citenamefont
  {Hirose}(1999)}]{ono1999timesaving}%
  \BibitemOpen
  \bibfield  {author} {\bibinfo {author} {\bibfnamefont {T.}~\bibnamefont
  {Ono}}\ and\ \bibinfo {author} {\bibfnamefont {K.}~\bibnamefont {Hirose}},\
  }\bibfield  {title} {\enquote {\bibinfo {title} {Timesaving double-grid
  method for real-space electronic-structure calculations},}\ }\href
  {https://doi.org/10.1103/PhysRevLett.82.5016} {\bibfield  {journal} {\bibinfo
   {journal} {Phys. Rev. Lett.}\ }\textbf {\bibinfo {volume} {82}},\ \bibinfo
  {pages} {5016} (\bibinfo {year} {1999})}\BibitemShut {NoStop}%
\bibitem [{\citenamefont {Ono}\ and\ \citenamefont
  {Hirose}(2005)}]{ono2005real}%
  \BibitemOpen
  \bibfield  {author} {\bibinfo {author} {\bibfnamefont {T.}~\bibnamefont
  {Ono}}\ and\ \bibinfo {author} {\bibfnamefont {K.}~\bibnamefont {Hirose}},\
  }\bibfield  {title} {\enquote {\bibinfo {title} {Real-space
  electronic-structure calculations with a time-saving double-grid
  technique},}\ }\href {https://doi.org/10.1103/PhysRevB.72.085115} {\bibfield
  {journal} {\bibinfo  {journal} {Phys. Rev. B}\ }\textbf {\bibinfo {volume}
  {72}},\ \bibinfo {pages} {085115} (\bibinfo {year} {2005})}\BibitemShut
  {NoStop}%
\bibitem [{\citenamefont {Tsukamoto}\ \emph {et~al.}(2013)\citenamefont
  {Tsukamoto}, \citenamefont {Caciuc}, \citenamefont {Atodiresei},\ and\
  \citenamefont {Bl{\"u}gel}}]{tsukamoto2013tuning}%
  \BibitemOpen
  \bibfield  {author} {\bibinfo {author} {\bibfnamefont {S.}~\bibnamefont
  {Tsukamoto}}, \bibinfo {author} {\bibfnamefont {V.}~\bibnamefont {Caciuc}},
  \bibinfo {author} {\bibfnamefont {N.}~\bibnamefont {Atodiresei}},\ and\
  \bibinfo {author} {\bibfnamefont {S.}~\bibnamefont {Bl{\"u}gel}},\ }\bibfield
   {title} {\enquote {\bibinfo {title} {Tuning electron transport through
  molecular junctions by chemical modification of the molecular core:
  First-principles study},}\ }\href
  {https://doi.org/10.1103/PhysRevB.88.125436} {\bibfield  {journal} {\bibinfo
  {journal} {Phys. Rev. B}\ }\textbf {\bibinfo {volume} {88}},\ \bibinfo
  {pages} {125436} (\bibinfo {year} {2013})}\BibitemShut {NoStop}%
\bibitem [{\citenamefont {Vosko}, \citenamefont {Wilk},\ and\ \citenamefont
  {Nusair}(1980)}]{vosko1980accurate}%
  \BibitemOpen
  \bibfield  {author} {\bibinfo {author} {\bibfnamefont {S.~H.}\ \bibnamefont
  {Vosko}}, \bibinfo {author} {\bibfnamefont {L.}~\bibnamefont {Wilk}},\ and\
  \bibinfo {author} {\bibfnamefont {M.}~\bibnamefont {Nusair}},\ }\bibfield
  {title} {\enquote {\bibinfo {title} {Accurate spin-dependent electron liquid
  correlation energies for local spin density calculations: a critical
  analysis},}\ }\href {https://doi.org/10.1139/p80-159} {\bibfield  {journal}
  {\bibinfo  {journal} {Canadian Journal of physics}\ }\textbf {\bibinfo
  {volume} {58}},\ \bibinfo {pages} {1200--1211} (\bibinfo {year}
  {1980})}\BibitemShut {NoStop}%
\bibitem [{\citenamefont {Troullier}\ and\ \citenamefont
  {Martins}(1991)}]{troullier1991efficient}%
  \BibitemOpen
  \bibfield  {author} {\bibinfo {author} {\bibfnamefont {N.}~\bibnamefont
  {Troullier}}\ and\ \bibinfo {author} {\bibfnamefont {J.~L.}\ \bibnamefont
  {Martins}},\ }\bibfield  {title} {\enquote {\bibinfo {title} {Efficient
  pseudopotentials for plane-wave calculations. ii. operators for fast
  iterative diagonalization},}\ }\href
  {https://doi.org/10.1103/PhysRevB.43.8861} {\bibfield  {journal} {\bibinfo
  {journal} {Phys. Rev. B}\ }\textbf {\bibinfo {volume} {43}},\ \bibinfo
  {pages} {8861} (\bibinfo {year} {1991})}\BibitemShut {NoStop}%
\bibitem [{NCP()}]{NCPP}%
  \BibitemOpen
  \href@noop {} {}\bibinfo {note} {We used the norm-conserving pseudopotentials
  NCPS97 constructed by K. Kobayashi. See K. Kobayashi, Comput. Mater. Sci.,
  14, 72 (1999).}\BibitemShut {Stop}%
\bibitem [{\citenamefont {Chelikowsky}, \citenamefont {Troullier},\ and\
  \citenamefont {Saad}(1994)}]{chelikowsky1994finite}%
  \BibitemOpen
  \bibfield  {author} {\bibinfo {author} {\bibfnamefont {J.~R.}\ \bibnamefont
  {Chelikowsky}}, \bibinfo {author} {\bibfnamefont {N.}~\bibnamefont
  {Troullier}},\ and\ \bibinfo {author} {\bibfnamefont {Y.}~\bibnamefont
  {Saad}},\ }\bibfield  {title} {\enquote {\bibinfo {title}
  {Finite-difference-pseudopotential method: Electronic structure calculations
  without a basis},}\ }\href {https://doi.org/10.1103/PhysRevLett.72.1240}
  {\bibfield  {journal} {\bibinfo  {journal} {Phys. Rev. Lett.}\ }\textbf
  {\bibinfo {volume} {72}},\ \bibinfo {pages} {1240} (\bibinfo {year}
  {1994})}\BibitemShut {NoStop}%
\bibitem [{\citenamefont {Chelikowsky}\ \emph {et~al.}(1994)\citenamefont
  {Chelikowsky}, \citenamefont {Troullier}, \citenamefont {Wu},\ and\
  \citenamefont {Saad}}]{chelikowsky1994higher}%
  \BibitemOpen
  \bibfield  {author} {\bibinfo {author} {\bibfnamefont {J.~R.}\ \bibnamefont
  {Chelikowsky}}, \bibinfo {author} {\bibfnamefont {N.}~\bibnamefont
  {Troullier}}, \bibinfo {author} {\bibfnamefont {K.}~\bibnamefont {Wu}},\ and\
  \bibinfo {author} {\bibfnamefont {Y.}~\bibnamefont {Saad}},\ }\bibfield
  {title} {\enquote {\bibinfo {title} {Higher-order finite-difference
  pseudopotential method: An application to diatomic molecules},}\ }\href
  {https://doi.org/10.1103/PhysRevB.50.11355} {\bibfield  {journal} {\bibinfo
  {journal} {Phys. Rev. B}\ }\textbf {\bibinfo {volume} {50}},\ \bibinfo
  {pages} {11355} (\bibinfo {year} {1994})}\BibitemShut {NoStop}%
\bibitem [{\citenamefont {Yoshida}\ \emph {et~al.}(2022)\citenamefont
  {Yoshida}, \citenamefont {Tanaka}, \citenamefont {Ataka}, \citenamefont
  {Hoshina},\ and\ \citenamefont {Furuya}}]{yoshida2022field}%
  \BibitemOpen
  \bibfield  {author} {\bibinfo {author} {\bibfnamefont {C.}~\bibnamefont
  {Yoshida}}, \bibinfo {author} {\bibfnamefont {T.}~\bibnamefont {Tanaka}},
  \bibinfo {author} {\bibfnamefont {T.}~\bibnamefont {Ataka}}, \bibinfo
  {author} {\bibfnamefont {M.}~\bibnamefont {Hoshina}},\ and\ \bibinfo {author}
  {\bibfnamefont {A.}~\bibnamefont {Furuya}},\ }\bibfield  {title} {\enquote
  {\bibinfo {title} {Field-free reliable magnetization switching in a
  three-terminal perpendicular magnetic tunnel junction via spin-orbit torque,
  spin-transfer torque, and voltage-controlled magnetic anisotropy},}\ }\href
  {https://doi.org/10.1088/1361-6463/ac791f} {\bibfield  {journal} {\bibinfo
  {journal} {Journal of Physics D: Applied Physics}\ }\textbf {\bibinfo
  {volume} {55}},\ \bibinfo {pages} {365003} (\bibinfo {year}
  {2022})}\BibitemShut {NoStop}%
\bibitem [{\citenamefont {Butler}\ \emph {et~al.}(2001)\citenamefont {Butler},
  \citenamefont {Zhang}, \citenamefont {Schulthess},\ and\ \citenamefont
  {MacLaren}}]{butler2001spin}%
  \BibitemOpen
  \bibfield  {author} {\bibinfo {author} {\bibfnamefont {W.~H.}\ \bibnamefont
  {Butler}}, \bibinfo {author} {\bibfnamefont {X.-G.}\ \bibnamefont {Zhang}},
  \bibinfo {author} {\bibfnamefont {T.~C.}\ \bibnamefont {Schulthess}},\ and\
  \bibinfo {author} {\bibfnamefont {J.~M.}\ \bibnamefont {MacLaren}},\
  }\bibfield  {title} {\enquote {\bibinfo {title} {Spin-dependent tunneling
  conductance of $\mathrm{Fe}|\mathrm{MgO}|\mathrm{Fe}$ sandwiches},}\ }\href
  {https://doi.org/10.1103/PhysRevB.63.054416} {\bibfield  {journal} {\bibinfo
  {journal} {Phys. Rev. B}\ }\textbf {\bibinfo {volume} {63}},\ \bibinfo
  {pages} {054416} (\bibinfo {year} {2001})}\BibitemShut {NoStop}%
\bibitem [{\citenamefont {Mathon}\ and\ \citenamefont
  {Umerski}(2001)}]{mathon2001theory}%
  \BibitemOpen
  \bibfield  {author} {\bibinfo {author} {\bibfnamefont {J.}~\bibnamefont
  {Mathon}}\ and\ \bibinfo {author} {\bibfnamefont {A.}~\bibnamefont
  {Umerski}},\ }\bibfield  {title} {\enquote {\bibinfo {title} {Theory of
  tunneling magnetoresistance of an epitaxial fe/mgo/fe(001) junction},}\
  }\href {https://doi.org/10.1103/PhysRevB.63.220403} {\bibfield  {journal}
  {\bibinfo  {journal} {Phys. Rev. B}\ }\textbf {\bibinfo {volume} {63}},\
  \bibinfo {pages} {220403} (\bibinfo {year} {2001})}\BibitemShut {NoStop}%
\bibitem [{\citenamefont {Masuda}\ \emph {et~al.}(2021)\citenamefont {Masuda},
  \citenamefont {Itoh}, \citenamefont {Sonobe}, \citenamefont {Sukegawa},
  \citenamefont {Mitani},\ and\ \citenamefont {Miura}}]{masuda2021interfacial}%
  \BibitemOpen
  \bibfield  {author} {\bibinfo {author} {\bibfnamefont {K.}~\bibnamefont
  {Masuda}}, \bibinfo {author} {\bibfnamefont {H.}~\bibnamefont {Itoh}},
  \bibinfo {author} {\bibfnamefont {Y.}~\bibnamefont {Sonobe}}, \bibinfo
  {author} {\bibfnamefont {H.}~\bibnamefont {Sukegawa}}, \bibinfo {author}
  {\bibfnamefont {S.}~\bibnamefont {Mitani}},\ and\ \bibinfo {author}
  {\bibfnamefont {Y.}~\bibnamefont {Miura}},\ }\bibfield  {title} {\enquote
  {\bibinfo {title} {Interfacial giant tunnel magnetoresistance and
  bulk-induced large perpendicular magnetic anisotropy in (111)-oriented
  junctions with fcc ferromagnetic alloys: A first-principles study},}\ }\href
  {https://doi.org/10.1103/PhysRevB.103.064427} {\bibfield  {journal} {\bibinfo
   {journal} {Phys. Rev. B}\ }\textbf {\bibinfo {volume} {103}},\ \bibinfo
  {pages} {064427} (\bibinfo {year} {2021})}\BibitemShut {NoStop}%
\bibitem [{\citenamefont {Karpan}\ \emph {et~al.}(2008)\citenamefont {Karpan},
  \citenamefont {Khomyakov}, \citenamefont {Starikov}, \citenamefont
  {Giovannetti}, \citenamefont {Zwierzycki}, \citenamefont {Talanana},
  \citenamefont {Brocks}, \citenamefont {Van Den~Brink},\ and\ \citenamefont
  {Kelly}}]{karpan2008theoretical}%
  \BibitemOpen
  \bibfield  {author} {\bibinfo {author} {\bibfnamefont {V.}~\bibnamefont
  {Karpan}}, \bibinfo {author} {\bibfnamefont {P.}~\bibnamefont {Khomyakov}},
  \bibinfo {author} {\bibfnamefont {A.}~\bibnamefont {Starikov}}, \bibinfo
  {author} {\bibfnamefont {G.}~\bibnamefont {Giovannetti}}, \bibinfo {author}
  {\bibfnamefont {M.}~\bibnamefont {Zwierzycki}}, \bibinfo {author}
  {\bibfnamefont {M.}~\bibnamefont {Talanana}}, \bibinfo {author}
  {\bibfnamefont {G.}~\bibnamefont {Brocks}}, \bibinfo {author} {\bibfnamefont
  {J.}~\bibnamefont {Van Den~Brink}},\ and\ \bibinfo {author} {\bibfnamefont
  {P.~J.}\ \bibnamefont {Kelly}},\ }\bibfield  {title} {\enquote {\bibinfo
  {title} {Theoretical prediction of perfect spin filtering at interfaces
  between close-packed surfaces of ni or co and graphite or graphene},}\ }\href
  {https://doi.org/10.1103/PhysRevB.80.035408} {\bibfield  {journal} {\bibinfo
  {journal} {Phys. Rev. B}\ }\textbf {\bibinfo {volume} {78}},\ \bibinfo
  {pages} {195419} (\bibinfo {year} {2008})}\BibitemShut {NoStop}%
\bibitem [{\citenamefont {Li}\ \emph {et~al.}(2014)\citenamefont {Li},
  \citenamefont {Xue}, \citenamefont {Abru\~na},\ and\ \citenamefont
  {Ralph}}]{li2014magnetic}%
  \BibitemOpen
  \bibfield  {author} {\bibinfo {author} {\bibfnamefont {W.}~\bibnamefont
  {Li}}, \bibinfo {author} {\bibfnamefont {L.}~\bibnamefont {Xue}}, \bibinfo
  {author} {\bibfnamefont {H.~D.}\ \bibnamefont {Abru\~na}},\ and\ \bibinfo
  {author} {\bibfnamefont {D.~C.}\ \bibnamefont {Ralph}},\ }\bibfield  {title}
  {\enquote {\bibinfo {title} {Magnetic tunnel junctions with
  single-layer-graphene tunnel barriers},}\ }\href
  {https://doi.org/10.1103/PhysRevB.89.184418} {\bibfield  {journal} {\bibinfo
  {journal} {Phys. Rev. B}\ }\textbf {\bibinfo {volume} {89}},\ \bibinfo
  {pages} {184418} (\bibinfo {year} {2014})}\BibitemShut {NoStop}%
\bibitem [{\citenamefont {Momma}\ and\ \citenamefont
  {Izumi}(2008)}]{momma2008vesta}%
  \BibitemOpen
  \bibfield  {author} {\bibinfo {author} {\bibfnamefont {K.}~\bibnamefont
  {Momma}}\ and\ \bibinfo {author} {\bibfnamefont {F.}~\bibnamefont {Izumi}},\
  }\bibfield  {title} {\enquote {\bibinfo {title} {{VESTA: a three-dimensional
  visualization system for electronic and structural analysis}},}\ }\href
  {https://doi.org/10.1107/S0021889808012016} {\bibfield  {journal} {\bibinfo
  {journal} {J. Appl. Cryst.}\ }\textbf {\bibinfo {volume} {41}},\ \bibinfo
  {pages} {653--658} (\bibinfo {year} {2008})}\BibitemShut {NoStop}%
\end{thebibliography}%
\end{document}


\title{
 Supporting Information of \\
 ``First-principle study of spin transport property in $L1_0$-FePd/Graphene heterojunction.''
}

\author{Hayato Adachi}
\affiliation{Department of Electrical and Electronic Engineering, Graduate School of Engineering, Kobe University, 1-1 Rokkodai-cho, Nada-ku, Kobe 651-8501, Japan}

\author{Ryuusuke Endo}
\affiliation{Department of Electrical and Electronic Engineering, Graduate School of Engineering, Kobe University, 1-1 Rokkodai-cho, Nada-ku, Kobe 651-8501, Japan}

\author{Hikari Shinya}
\affiliation{
 Center for Spintronics Research Network (CSRN), University of Tokyo,
 7-3-1, Hongo, Bunkyo-ku, Tokyo 113-8656, Japan
}
\affiliation{
 Institute for Chemical Research, Kyoto University,
 Gokasho, Uji, Kyoto 611-0011, Japan
}
\affiliation{
 Center for Science and Innovation in Spintronics (CSIS), Tohoku University,
 2-1-1, Katahira, Aoba-ku, Miyagi 980-8577, Japan
}
\affiliation{
 Center for Spintronics Research Network (CSRN), Osaka University,
 1-3, Machikaneyama, Toyonaka, Osaka 560-8531, Japan
}

\author{Hiroshi Naganuma}
\affiliation{Center for Innovative Integrated Electronics Systems (CIES), Tohoku University, 468-1 Aramaki Aza Aoba, Aoba, Sendai, Miyagi, 980-8572, Japan}
\affiliation{Center for Spintronics Integrated Systems (CSIS), Tohoku University, 2-2-1 Katahira Aoba, Sendai, Miyagi 980-8577 Japan}
\affiliation{Center for Spintronics Research Network (CSRN), Tohoku University, 2-1-1 Katahira, Aoba, Sendai, Miyagi 980-8577 Japan}
\affiliation{Graduate School of Engineering, Tohoku University, 6-6-05, Aoba, Aoba-ku, Sendai, Miyagi, 980-8579, Japan}

\author{Tomoya Ono}
\affiliation{Department of Electrical and Electronic Engineering, Graduate School of Engineering, Kobe University, 1-1 Rokkodai-cho, Nada-ku, Kobe 651-8501, Japan}

\author{Mitsuharu Uemoto}
\email{uemoto@eedept.kobe-u.ac.jp}
\affiliation{Department of Electrical and Electronic Engineering, Graduate School of Engineering, Kobe University, 1-1 Rokkodai-cho, Nada-ku, Kobe 651-8501, Japan}

\date{\today}
\maketitle

\clearpage
\section*{S1.~computational conditions for calculating the density of states and band structure}

\begin{figure*}[htbp]
 \centering
 \includegraphics[width=0.75\textwidth]{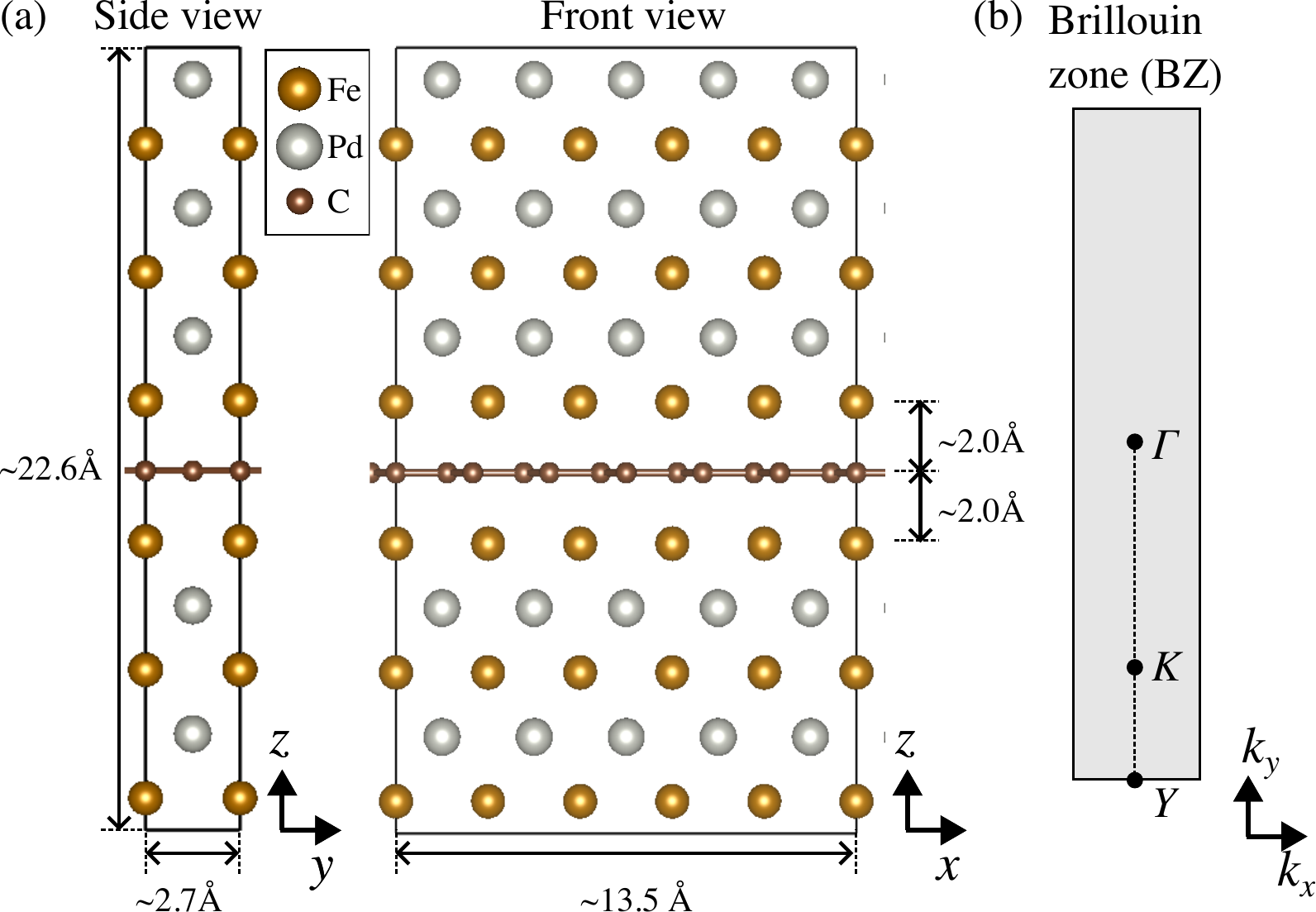}
 \caption{
 Computational model used for calculation of the density of states (DoS) and band structure.
 (a) Side view ($yz$ cross-section) and front view ($xz$ cross-section).
 (b) Brillouin zone with labeled K points and selected path for band plotting.
 }
 \label{fig:band_model}
\end{figure*}

The density of states (DoS) and band structure are calculated using the Vienna Ab-initio Software Package (VASP) code, which is based on density functional theory with plane-wave expansion and the projector augmented-wave method.
The computational model is extracted from the scattering region of 'FePd(001)/1-Gr/FePd(001)' heterojunction. [See Fig.~\ref{fig:band_model}.]
We utilize the supercell of $13.5~\text{\AA} \times 2.7~\text{\AA} \times 22.6~\text{\AA}$ containing 67 atoms.
The plane wave basis cutoff energy is set to be 600~eV, and $k$-point sampling of the first Brillouin zone is set to be $2 \times 10 \times 1$ $\Gamma$-centered grids for the ground state calculation.
For the DoS calculation, a $3 \times 15 \times 1$ grid is also employed for $k$-space integration.
Besides, for the band structure calculation, we sampled 80 points on the $k_y$-segment trough $\Gamma$-$K$-$Y$ points, illustrated as the dotted line in Fig.~\ref{fig:band_model}(b).